\documentclass[12pt]{article}
\newcommand{\Comment}[1]{{}}
\usepackage[textwidth = 450 pt, textheight = 630 pt]{geometry}
\usepackage[parfill]{parskip}
\usepackage{amssymb,euscript,amsmath,amsfonts}
\usepackage{tikz,tikz-cd,url}
\usetikzlibrary{decorations.pathreplacing,calligraphy}
\usepackage{xcolor}
\usepackage{graphicx}
\usepackage{color}
\usepackage{bbm}
\usepackage{tensor}
\usepackage{slashed}
\newcommand{\inv}[1]{\frac{1}{#1}}
\newcommand{\brac}[1]{ \left( #1 \right)}
\renewcommand{\cal}[1]{\mathcal{#1}}
\newcommand{\calz}{\cal{Z}}
\newcommand{\bcalz}{\bar{\cal{Z}}}
\newcommand{\hcalz}{\hat{\calz}}
\newcommand{\hbcalz}{\hat{\bcalz}}
\newcommand{\bpsi}{\bar{\psi}}
\newcommand{\hpsi}{\hat{\psi}}
\newcommand{\hbpsi}{\hat{\bpsi}}
\newcommand{\hd}{\hat{D}}
\newcommand{\hbd}{\hat{\bar{D}}}
\newcommand{\pf}{\partial f}
\newcommand{\mn}{\mu\nu}
\newcommand{\ch}{\cal{H}}
\newcommand{\sbrac}[1]{\left[ #1 \right]}
\newcommand{\bpf}{\bar{\partial} \bar{f}}
\newcommand{\bb}[1]{\mathbb{#1}}
\renewcommand{\frak}[1]{\mathfrak{#1}}
\newcommand{\bbm}[1]{\mathbbm{#1}}

\definecolor{MyDarkBlue}{rgb}{0.15,0.15,0.45}
\usepackage[linktocpage=true]{hyperref}
\hypersetup{
colorlinks=true,
citecolor=MyDarkBlue,
linkcolor=MyDarkBlue,
urlcolor=MyDarkBlue,
pdfauthor={Neil Lambert },
pdftitle={ },
pdfsubject={hep-th}
}

\usepackage{hypernat}
\usepackage{physics}
\usepackage{accents}
\usepackage{bm}
\usepackage[sorting=none]{biblatex}
\begin{document}

   \vspace{1.8truecm}

 \centerline{\Huge  {  Non-Relativistic M2-Branes and the }} \vskip 12pt
 \centerline{\Huge  {  AdS/CFT Correspondence}}

\centerline{\LARGE \bf {\sc  }} \vspace{2truecm} \thispagestyle{empty} \centerline{
    {\large {{\sc Neil~Lambert}}}\footnote{E-mail address: \href{mailto:neil.lambert@kcl.ac.uk}{\tt neil.lambert@kcl.ac.uk} } and  {\large {{\sc Joseph~Smith}}}\footnote{E-mail address: \href{mailto:joseph.m.smith@kcl.ac.uk}{\tt joseph.m.smith@kcl.ac.uk} }
  }

\vspace{1cm}
\centerline{{\it Department of Mathematics}}
\centerline{{\it King's College London }} 
\centerline{{\it The Strand }} 
\centerline{{\it  WC2R 2LS, UK}} 

\vspace{1.0truecm}

\thispagestyle{empty}

\centerline{\sc Abstract}
\vspace{0.4truecm}
\begin{center}
\begin{minipage}[c]{360pt}{
    \noindent}

A non-relativistic limit of the AdS/CFT correspondence is studied in the context of M2-branes. On the field theory side this corresponds to a near-BPS limit of ABJM that localises onto solutions of Hitchin's equations. It is shown that the symmetries of the theory include an infinite-dimensional enhancement of the spatial symmetry algebra corresponding to time-dependent holomorphic transformations. Taking the limit of the gravitational dual splits the geometry into three 'large' directions and eight 'small' directions and corresponds to the Membrane-Newton-Cartan limit of eleven-dimensional supergravity. This has the effect of reducing the $AdS_4$ factor to an $AdS_2$ factor for the near-horizon limit of the M2-brane metric. Evidence is presented  that the duality is maintained after the limit.

\end{minipage}
\end{center}

\newpage
\tableofcontents

\section{Introduction}\label{sect: Intro}

There has been a recent Renaissance in the study of non-Lorentzian physics and its applications to relativistic theories. A topic that has garnered considerable attention is non-relativistic limits of string theory. While this area has been studied for a long time \cite{Gomis:2000bd, Gomis:2005pg}, its development has recently been spurred by a deeper understanding of the non-relativistic worldsheet theory and the spacetime geometry to which it couples \cite{Bergshoeff:2019pij, Bidussi:2021ujm, Oling:2022fft}. The key point (first put forward in \cite{Andringa:2012uz}) is that strings naturally couple to non-Lorentzian manifolds with two distinguished directions, known as String Newton-Cartan (SNC) geometries, in contrast to the single direction in a standard Newton-Cartan geometry. In practice this means that the relativistic metric is split into two pieces, $\tau_{\mn}$ and $h^{\mn}$, where $\tau_{\mn}$ has two non-zero eigenvalues (one positive and one negative) and $h^{\mn}$ has eight positive eigenvalues\footnote{Note that we are assuming we work in the critical dimension $D=10$.}. As the low energy dynamics of a relativistic string theory are governed by a supergravity theory a corresponding non-relativistic limit of this can be taken after decomposing the metric into the corresponding SNC structures \cite{Bergshoeff:2021bmc, Bergshoeff:2023ogz}. As in the relativistic case, we can view non-relativistic string theory as the UV completion of the corresponding non-relativistic supergravity theory. Since string theory contains more extended objects than just the fundamental string, it is natural to extend this idea to more general $p$-brane geometries \cite{Bergshoeff:2023rkk}\footnote{Limits of $p$-brane geometries can also be studied in the framework of SNC geometries \cite{Avila:2023aey}.}. In \cite{Blair:2021waq} this was applied to the case of M2-branes in eleven-dimensional supergravity, where the non-relativistic limit was taken and found to give a gravitational theory for Membrane Newton-Cartan (MNC) geometries with three distinguished directions. One may hope that, as for non-relativistic limits of the ten-dimensional supergravity theories, there is a well-defined non-relativistic M-theory that serves as this theory's UV completion\footnote{See \cite{Roychowdhury:2022est} for a discussion of the MNC limit of the M2 worldvolume action.}.

Another active area of research is the process of obtaining non-Lorentzian quantum field theories from their Lorentzian counterparts. This includes directly taking limits of our coordinates and fields, as well as more exotic methods such as null reductions and variations thereof (see \cite{Baiguera:2023fus} for a recent review). Our method of interest will be taking a Galilean $c\to\infty$ limit. Typically the field theories  studied in this way are massive, and there is a well-defined way of finding a non-relativistic limit. In contrast, much less emphasis has been placed on finding limits of conformal field theories (CFTs); indeed, a naive non-relativistic limit of a massless field gives a theory with trivial dynamics. The question of finding an interesting limit for the ABJM theory was tackled in \cite{Lambert:2019nti}, where a scaling limit was found that gives a non-relativistic theory with the same amount of supersymmetry as the parent theory. The dynamics of the theory are non-trivial and, in the simplest case, can be identified with motion on the moduli space of Hitchin's equations \cite{Hitchin:1986vp}. While an interesting field theory, there are still open questions as to its structure. For example, the symmetries of the theory after the limit are unknown. Also,
preservation of supersymmetry requires an ad-hoc field redefinition to be performed before the limit is taken; it would be beneficial to have an argument as to what this shift signifies in order to understand the limit's physical interpretation.

Given that non-relativistic limits can be taken of both field theories and gravity, it is interesting to ask whether these ideas can be applied to the AdS/CFT correspondence: in other words, can non-relativistic limits be taken on both sides such that the duality is maintained? This has previously been studied in the context of the duality between four-dimensional $\cal{N}=4$ super Yang-Mills and type IIB string theory on $AdS_5\times S^5$ by taking a decoupling limit of the field theory that isolates operators near a BPS bound. These are known as Spin Matrix Theory limits \cite{Harmark:2014mpa, Harmark:2017rpg, Baiguera:2022pll}; since the quantum numbers of the surviving operators satisfy certain relations the holographic dictionary can be used to translate these into a corresponding non-relativistic limit of the gravitational background. 

Our aim is to approach this question from a different perspective for the duality between the ABJM theory and M-theory on $AdS_4\times S^7/\bb{Z}_k$ \cite{Aharony:2008ug}. We have already discussed the non-trivial scaling limit of ABJM found in \cite{Lambert:2019nti}. As ABJM describes the low-energy dynamics of a stack of M2-branes on a $\bb{C}^4/\bb{Z}_k$ background, we can reinterpret the scaling as a non-relativistic limit of the M2-brane spacetime. This turns out to be exactly of the form required by the MNC limit of eleven-dimensional supergravity in \cite{Blair:2021waq}. As the limit leads to well-defined theories on both sides there is the potential that the duality is retained. We claim that this is indeed the case, and our goal in this work is to put forward evidence in favour of this conclusion.

%%%%%%%%%%%%%%%%%%%%%%%%%%%%%%%%%%%%%%%%%%%%%%%%%%%
There is a considerable body of work dedicated to finding non-relativistic AdS/CFT pairs by considering solutions of relativistic gravity theories with a Lifshitz scaling symmetry 
\cite{Son:2008ye,Balasubramanian:2008dm,Herzog:2008wg,Barbon:2008bg} or compactified directions that are either null or become null near the boundary \cite{Maldacena:2008wh,Dorey:2022cfn,Mouland:2023gcp,Goldberger:2008vg}. Unlike these works, the gravity theory we propose as the field theory's dual is inherently non-relativistic. This appears to manifest itself in the structure of symmetries of the theory: while the field theories obtained in the approaches discussed above possess Schr\"odinger symmetry, we will find something more exotic.
%%%%%%%%%%%%%%%%%%%%%%%%%%%%%%%%%%%%%%%%%%%%%%%%%%%

This paper is organised as follows. In section \ref{sect: toy} we take a naive non-relativistic limit of a Chern-Simons matter theory, which leads to a theory in which all symmetries can be given arbitrary time-dependence and we have no dynamics. In section \ref{sect: near bps limit} we apply a similar procedure to the ABJM theory, where we retain dynamics by ensuring that the equations defining half-BPS solutions are unchanged by the limit. We analyse the symmetries of the resulting theory and find that the spatial symmetries are enhanced to the two-dimensional Euclidean conformal algebra with arbitrary time-dependence. These should be interpreted as redundancies, as evidenced by the corresponding conserved charge reducing to a boundary term. The analogous limit of the gravitational dual is studied in section \ref{sect: 11d gravity}, where the appropriate non-relativistic theory is the membrane Newton-Cartan limit of eleven-dimensional supergravity found in \cite{Blair:2021waq}. We show that the physical symmetries of the field theory are realised in the gravitational solution, leading us to propose that the duality between the two is maintained after taking the non-relativistic limit on both sides. Our findings are summarised in section \ref{sect: conclusion} and avenues for further work are discussed. We also include supplementary material as appendices. The analysis of the fermionic terms in the field is performed in appendix \ref{sect: fermions}, and the equations of motion for the field theory are collected in appendix \ref{sect: eom}. In appendix \ref{sect: orbifold} we discuss the non-relativistic limit of orbifold geometries.

\section{Non-Relativistic Limits of Three-Dimensional Chern-Simons-Matter Theories}\label{sect: toy}
 Let us consider for illustration a Bosonic Chern-Simons matter theory with action of the form
\begin{align}
S = - \frac{1}{c} \tr \int d^3x  \sqrt{-g} \brac{ g^{\mu\nu}  D_\mu \hcalz^M  D_\nu \hbcalz_M  + \hat{V}(\hcalz^M,\hbcalz_M) } + S_{CS}\ ,
\end{align}
where
\begin{align}
 S_{CS}= \frac{k}{4\pi }\tr\int 	 d^3x \, \varepsilon^{\mu\nu\lambda } \brac{A_\mu\partial _\nu A_\lambda - \frac{2i}{3}A_\mu A_\nu A_\lambda } \ ,
\end{align}
is a Chern-Simons term and $\hat{V}$ a potential. Here $c$ is identified with the speed of light and we work with the Minkowski metric $g_{\mn} = {\rm diag}(-c^2,1,1)$. 
%%%%%%%%%%%%%%%%%%%%%%%
We will consider a theory for which the scalars are in the adjoint of $U(N)$,
%%%%%%%%%%%%%%%%%%%%%%%%%
so $D_\mu \hcalz^M = \partial_\mu \hcalz^M - i [A_\mu, \hcalz^M]$,
%%%%%%%%%%%%%%%%%%%%%%%%%
with $M$ a flavour index.
%%%%%%%%%%%%%%%%%%%%%%%%%

As written this is a relativistic field theory. However if the potential contains an explicit mass term
\begin{align}
	\hat{V}(\hcalz^M,\hbcalz_M)  = m^2c^2  \hcalz^M\hbcalz_M + V(\hcalz^M,\hbcalz_M)\ ,
\end{align}
then we can consider a non-relativistic limit by writing
\begin{align}
\hcalz^M = e^{-imc^2 t}	 \calz^M\ .
\end{align}
In the $c\to\infty$ limit the action becomes
\begin{align}
S = \tr \int  dt d^2x  \brac{i m \brac{ \calz^M D_t \bcalz_M -   D_t \calz^M  \bcalz_M} - D_i\calz^M D_i\bcalz_M - V(\calz^M, \bcalz^M)} + S_{CS} \ .
\end{align}
The resulting equations of motion become that of a Schr\"odinger-type theory coupled to a non-Abelian gauge field.

On the other hand if the action is classically invariant under the scale transformations
\begin{align}
x^\mu& \to \lambda x^\mu\nonumber\\
A_\mu &\to \lambda^{-1}A_\mu\nonumber\\
\hat \calz^M& \to \lambda^{-\tfrac12} \hat \calz^M\ ,\end{align}
then there is no mass term and we can't take such a non-relativistic limit. 
Instead we would like to consider the following rescaling of the spacetime metric (in complex coordinates $z= x^1+ix^2$ and setting $c=1$):
\begin{align} \label{eq: rescaled metric}
g_{\mu\nu} = \left(\begin{matrix}-1&0&0\\0&0&\frac{\omega^2}{2}\\0&\frac{\omega^2}{2} &0\end{matrix}\right)	\ ,
\end{align}
which can also be viewed as a rescaling of the spatial coordinates. Since we are in a conformal field theory this is the same as rescaling time $x^0\to \omega^{-1} x^0$ and leaving space unchanged (as well as a suitable action on the fields). We are interested in the limit $\omega \to 0 $ which, in terms of $x^0=ct$, is conformally equivalent to a non-relativistic limit $c\to \infty$ (but would also include a rescaling of the scalars fields). 

We note that 
\begin{align}
\sqrt{-g}g^{\mu\nu} = {\omega^2}\left(\begin{matrix}-1&0&0\\0&0&\frac{2}{\omega^2}\\0&\frac{2}{\omega^2} &0\end{matrix}\right)	 = \left(\begin{matrix}-{\omega^2}&0&0\\0&0&2\\0& 2&0\end{matrix}\right)\ .
\end{align}
Thus the limit $\omega\to 0$ is smooth and the action reduces to
\begin{align}
S = - 2 \tr \int d^3x \brac{ D \calz^M \bar D \bcalz_M + D \bcalz^M \bar D \calz_M }   + S_{CS}\ ,
\end{align}
where $D=D_z$, $\bar D=D_{\bar z}$ and we trivially identify ${\calz}^M=\hat \calz^M$. Note that the Chern-Simons term is unaffected by this deformation of the metric but the potential term vanishes as $\sqrt{-g}\to 0$. 

The equations of motion are now 
\begin{align}
F_{z\bar z} & = 0\nonumber\\
F_{0z} & = \frac{2\pi}{k}\left( Z^M D \bar Z^M - DZ^M  \bar Z^M  \right)   \nonumber \\
(D\bar D + \bar D D )\calz^M	 & = 0\ .
\end{align}
The first equation tells us that the spatial gauge field is pure gauge. 
%%%%%%%%%%%%%%%%%%%%%%%%%%%%%%%%%
As we are working on a flat manifold there are no topological subtleties 
%%%%%%%%%%%%%%%%%%%%%%%%%%%%%%%%%
and we can simply take $A_z =A_{\bar z}=0$. The equations are now just
\begin{align}
-\partial A_0 &= 	\frac{2\pi}{k}\left( \calz^M \partial \bcalz^M -  \partial \calz^M    \bcalz^M\right)\nonumber\\
\partial\bar\partial \calz^M & =0  
\end{align}
A natural class of solutions consists of setting $\bar \partial   \calz^M=0$ and hence 
\begin{align}\label{A0sol}
A_0  = \frac{2\pi}{k} \calz^M \bcalz^M\ .
\end{align}
(There is a similar class where $ \partial   Z^M=0$.)
Note that the time dependence has played no role.   

More generally this action is invariant under an infinite dimensional group of diffeomorphisms and Weyl transformations. However in general these are not symmetries  but redundancies in the description.  Rather, for a given metric $g_{\mu\nu}$, the only transformations that lead to symmetries\footnote{We will ignore global symmetries for the time being.} are those for which
\begin{align}\label{Syms}
\frac{\partial x^\lambda}{\partial x'^\mu }\frac{\partial x^\rho}{\partial x'^\nu }g_{\lambda\rho}(x) = \Omega^2 g_{\mu\nu}(x')\ , 	
\end{align}
for some $\Omega$. For the standard choice of $g=\eta$ this leads to the conformal group  $SO(2,3)$; this remains true for the rescaled metric \eqref{eq: rescaled metric} when $\omega$ is non-zero, as a field redefinition will set $\omega=1$. However, for $\omega=0$ the metric is degenerate and we must be more careful. From the action we see the quantity of interest is $\sqrt{-g}g^{\mu\nu}$, which must have the transformation
\begin{align}
	\sqrt{-g}g^{\mu\nu }{\det\left(\frac{\partial x}{\partial x'}\right)}\frac{\partial x'^\lambda}{\partial x^\mu }\frac{\partial x'^\rho}{\partial x^\nu } = \sqrt{-g}\Omega g^{\lambda\rho} \ ,
	%= \Omega \left(\begin{matrix}0&0&0\\0&0&1\\0&1 &0\end{matrix}\right)	
\end{align}
with
\begin{align}
\sqrt{-g}g^{\mu\nu }= \left(\begin{matrix}0&0&0\\0&0&1\\0&1 &0\end{matrix}\right)	\ .
\end{align}
Looking at the timelike components we   see that $x'^0$ can be any function of $x^0$ but cannot depend on $z$ or  $\bar z$.
The other components restrict $z'$ to holomorphic functions of $z$ but arbitrary functions of $x^0$ with 
\begin{align}
\Omega = 	{\det\left(\frac{\partial x}{\partial x'}\right)}\frac{\partial z'}{\partial z}\frac{\partial \bar z'}{\partial \bar z} =  \frac{\partial x^0}{\partial x'^0}\ .\end{align}
 
Thus, in contrast to the case with an invertible metric, the symmetry group is infinite dimensional.  Indeed it contains the infinite dimensional two-dimensional conformal group  consisting of holomorphic transformations (with time-dependent transformations) along with one-dimensional diffeomorphisms of time.  This is consistent with the solutions to the equations of motion that we found above.

We have seen that the dynamics of the theory are  totally unconstrained. Indeed, there is no notion of time.  Thus we have arrived at something we can think of as a non-relativistic topological gauge theory. 
Although curious,  without any dynamics this is of limited physical interest. We can perform a similar scaling but also rescale one of the scalar fields, say $\hat \calz^1= \omega^{-1} \calz^1$. This has the effect that the $\calz^1$ kinetic term remains non-zero and hence we retain some dynamics. This limit is more complicated than the one we have just discussed as it introduces divergent terms but these can be eliminated. We turn our attention to this construction in the next section for the specific case of M2-branes. As we shall soon see  we still find an infinite number of symmetries.

\section{Near-BPS Limit of ABJM} \label{sect: near bps limit}
\subsection{Scaling Limit} \label{sect: scaling}

Let us review the field theory obtained in \cite{Lambert:2019nti} using the approach of \cite{Mouland:2019zjr}. Our goal is to find a scaling limit of the ABJM theory that 'zooms in' on a class of $\inv{2}$-BPS solutions such that the dynamics of the theory reduces to the Manton approximation of geodesic motion on the moduli space of solutions \cite{Manton:1981mp}. This is achieved by finding a rescaling of both fields and coordinates under which the BPS equations are invariant and kinetic terms are suppressed relative to gradient terms. We will only focus on the Bosonic fields in the main body of the text, with the Fermions discussed in appendix \ref{sect: fermions}.

The ABJM theory \cite{Aharony:2008ug} is a 3d $\cal{N}=6$ superconformal $U(N) \times U(N)$ Chern-Simons matter theory with two $U(N)$ gauge fields $A^{L/R}$ and four complex scalar fields $\hat{\calz}^M$, where $M\in\{1,2,3,4\}$ is the $SU(4)$ R-symmetry index, in the bifundamental of $U(N) \times U(N)$ as its Bosonic field content. As in section \ref{sect: toy} we will work with Hermitian gauge fields throughout, so the covariant derivative of the scalar fields is given by
\begin{equation}
    \hat{D}_{\mu} \hat{\calz}^M = \hat{\partial}_{\mu} \hat{\calz}^M - i \hat{A}_{\mu}^L \hat{\calz}^M + i \hat{\calz}^M \hat{A}_{\mu}^R \ .
\end{equation}
The action for these fields is
\begin{align} \nonumber
    \hat{S}_B = \tr \int d^3\hat{x} \Bigg[& - \hat{D}_{\mu} \hat{\bcalz}_M \hat{D}^{\mu} \hat{\calz}^M + \frac{k \varepsilon^{\mn\rho}}{4\pi} \bigg( \hat{A}_{\mu}^L \hat{\partial}_{\nu} \hat{A}_{\rho}^L - \frac{2i}{3} \hat{A}_{\mu}^L \hat{A}_{\nu}^L \hat{A}_{\rho}^L - \hat{A}_{\mu}^R \hat{\partial}_{\nu} \hat{A}_{\rho}^R \\
    &+ \frac{2i}{3} \hat{A}_{\mu}^R \hat{A}_{\nu}^R \hat{A}_{\rho}^R \bigg) - \frac{8\pi^2}{3k^2 } \hat{\bar{\Upsilon}}^{K}_{IJ} \hat{\Upsilon}^{IJ}_{K} \Bigg] \ ,
\end{align}
where we have defined
\begin{equation}
    \hat{\Upsilon}^{MN}_{P} = [\hat{\calz}^M , \hat{\calz}^N ; \hat{\bcalz}_P] - \inv{2} \delta^M_P [ \hat{\calz}^Q , \hat{\calz}^N; \hat{\bcalz}_Q] + \inv{2} \delta^N_P [ \hat{\calz}^Q , \hat{\calz}^M ; \hat{\bcalz}_Q]\ ,
\end{equation}
using the notation
\begin{equation}
    [\hat{\calz}^M , \hat{\calz}^N ; \hat{\bcalz}_P] = \hat{\calz}^M \hat{\bcalz}_P \hat{\calz}^N - \hat{\calz}^N \hat{\bcalz}_P \hat{\calz}^M \ .
\end{equation}
Field configurations preserve a supercharge if the supersymmetry transformation of the Fermions vanish for the spinor parameter associated with that supercharge. In the ABJM theory this corresponds to solving the equations
\begin{equation} \label{eq: BPS initial}
    0 = - \gamma^{\mu} \hat{D}_{\mu} \hat{\calz}^N \xi_{MN} + \frac{2\pi}{k} [\hat{\calz}^P, \hat{\calz}^Q; \hat{\bcalz}_M] \xi_{PQ} + \frac{2\pi}{k} [\hat{\calz}^N, \hat{\calz}^P; \hat{\bcalz}_Q] \xi_{MN} \ ,
\end{equation}
where $\xi_{MN}$ are a set of spinor parameters that are antisymmetric in the R-symmetry indices and satisfy the reality condition
\begin{equation}
    \bar{\xi}^{MN} \equiv (\xi_{MN})^* = \inv{2} \epsilon^{MNPQ} \xi_{PQ}
\end{equation}
in a specific basis of the gamma matrices discussed further in appendix \ref{sect: fermions}. As we are interested in $\inv{2}$-BPS solutions\footnote{{\it i.e.} solutions for which half the supercharges are preserved.} we impose the conditions
\begin{subequations}
\begin{align}
    i \gamma^0 \xi_{1A} &= \xi_{1A} \ , \\
    i \gamma^0 \xi_{AB} &= - \xi_{AB} \ ,
\end{align}
\end{subequations}
where the index is $A\in\{2,3,4\}$, on the spinor parameters to reduce the degrees of freedom by half. With this choice the equations \eqref{eq: BPS initial} are then \cite{Kim:2009ny}
\begin{subequations} \label{eq: undeformed BPS equations}
\begin{align}
    \hat{D}_i \hat{\calz}^A &= 0 \ , \\ \label{eq: holomorphic BPS equation}
    \hat{\bar{D}} \hat{\calz}^1 &= 0 \ , \\
    [\hat{\calz}^1 , \hat{\calz}^2 ; \hat{\bcalz}_2] &= [\hat{\calz}^1 , \hat{\calz}^3 ; \hat{\bcalz}_3] = [\hat{\calz}^1 , \hat{\calz}^4 ; \hat{\bcalz}_4] \ , \\
    \hat{D}_0 \hat{\calz}^1 &= \frac{2\pi i}{3k} [\hat{\calz}^1 , \hat{\calz}^A ; \hat{\bcalz}_A] \ , \\ \label{eq: undeformed ZA eq}
    \hat{D}_0 \hat{\calz}^A &= \frac{2\pi i}{k} [\hat{\calz}^1, \hat{\calz}^A ; \hat{\bcalz}_1] \ , \\
    [\hat{\calz}^1 , \hat{\calz}^A ; \hat{\bcalz}_B] &= 0 \; (A\neq B) \ , \\
    [\hat{\calz}^A , \hat{\calz}^B ; \hat{\bcalz}_C] &= 0 \ ,
\end{align}
\end{subequations}
which we must supplement with the Gauss's law constraints
\begin{subequations} \label{eq: Gauss law}
\begin{align} \nonumber 
    \frac{ik}{2\pi} \hat{F}^L_{12} &= \hat{\calz}^M \hat{D}_0 \hat{\bcalz}_M - \hat{D}_0 \hat{\calz}^M \hat{\bcalz}_M \\
    &=\frac{4\pi i}{3k} \brac{ 2 \hat{\calz}^A [ \hat{\bcalz}_1, \hat{\bcalz}_A ; \hat{\calz}^1] - [ \hat{\calz}^1, \hat{\calz}^A ; \hat{\bcalz}_1] \hat{\bcalz}_A - [\hat{\calz}^1, \hat{\calz}^A ; \hat{\bcalz}_A] \hat{\bcalz}_1} \ , \\ \nonumber
    \frac{ik}{2\pi} \hat{F}_{12}^R &= \hat{D}_0 \hat{\bcalz}_M \hat{\calz}^M - \hat{\bcalz}_M \hat{D}_0 \hat{\calz}^M \\
    &= \frac{4\pi i}{3k} \brac{ 2 [ \hat{\bcalz}_1, \hat{\bcalz}_A ; \hat{\calz}^1] \hat{\calz}^A - \hat{\bcalz}_A [ \hat{\calz}^1, \hat{\calz}^A ; \hat{\bcalz}_1] - \hat{\bcalz}_1 [ \hat{\calz}^1 , \hat{\calz}^A ; \hat{\bcalz}_A] } \ .
\end{align}
\end{subequations}
We see that singling out the index $M=1$ in the spinor condition has singled out the field $\hat{\calz}^1$ in the BPS equations.

With the BPS equations in hand, we can now discuss coordinate scalings that satisfy the criteria discussed above. As ABJM has a scaling symmetry under which all spacetime coordinates transform homogeneously, a scaling for which both spatial coordinates scale in the same way can be put into the form
\begin{equation} \label{eq: scaling 1}
    (\hat{t}, \hat{x}^i) = ( t , \omega x^i)\ ,
\end{equation}
for a specific choice of $\omega$. For our purposes the scaling limit will take $\omega \to 0$. If the BPS equations are to be invariant under the scaling, then it appears that we must take the fields to have the scaling
\begin{subequations}
\begin{align}
    \hat{\calz}^1(\hat{t},\hat{x}) &= \calz^1(t,x) \ , \\
    \hat{\calz}^A(\hat{t},\hat{x}) &= \calz^A(t,x) \ , \\
    \hat{A}_0(\hat{t}, \hat{x}) &= A_0(t,x) \ , \\
    \hat{A}_i(\hat{t},\hat{x}) &= \inv{\omega} A_i(t,x) \ ,
\end{align}
\end{subequations}
which coincides with the scaling limit discussed in section \ref{sect: toy}. However, this is inconsistent with invariance of the constraints \eqref{eq: Gauss law} and, as seen previously, results in a theory with trivial dynamics. The resolution of this is to shift the timelike components of the gauge field to
\begin{subequations} \label{eq: A0 shift}
\begin{align}
    A^L_0 &= \cal{A}^L_0 - \frac{2\pi}{k} \calz^1 \bcalz_1 \ , \\
    A^R_0 &= \cal{A}^R_0 - \frac{2\pi}{k} \bcalz_1 \calz^1 \ ,
\end{align}
\end{subequations}
as this shifts \eqref{eq: undeformed ZA eq} to
\begin{equation}
    0 = D_0 \calz^A - \frac{2\pi i}{k} [\calz^1, \calz^A ; \bcalz_1] = \cal{D}_0 \calz^A\ ,
\end{equation}
while leaving $D_0\calz^1 = \cal{D}_0\calz^1$. 
%%%%%%%%%%%%%%%%%%%%%%%%%%%%%%%%
We note that since the difference of $A^{L/R}_0$ and $\cal{A}^{L/R}_0$ is a field in the adjoint representation of $U(N)_{L/R}$, $\cal{A}_0^{L/R}$ is still a $U(N)_{L/R}$ connection.
%%%%%%%%%%%%%%%%%%%%%%%%%%%%%%%%
Using this, we find that the scaling
\begin{subequations} \label{eq: scaling 2}
\begin{align}
    \hat{\calz}^1(\hat{t},\hat{x}) &= \inv{\omega}\calz^1(t,x) \ , \\
    \hat{\calz}^A(\hat{t},\hat{x}) &= \calz^A(t,x) \ , \\
    \hat{\cal{A}}_0(\hat{t}, \hat{x}) &= \cal{A}_0(t,x) \ , \\
    \hat{A}_i(\hat{t},\hat{x}) &= \inv{\omega} A_i(t,x) \ ,
\end{align}
\end{subequations}
leaves both the BPS equations and Gauss's law invariant.

The application of the scaling limit to the ABJM action was performed in \cite{Lambert:2019nti}\footnote{The scaling there differs by an overall conformal transformation from that discussed here, but as this is a symmetry of ABJM the outcome is the same.}, and we shall briefly review it for the Bosonic sector of the theory here. Note that from here onwards we will use $A_0$ instead of $\cal{A}_0$ for the shifted gauge field to simplify our notation. At finite $\omega$ the action takes the form
\begin{subequations}
\begin{equation}
    \hat{S}_B = \inv{\omega^2} S_{B,-} + S_{B,0} + O(\omega^2) \ .
\end{equation}
The finite term is
\begin{align} \nonumber
    S_{B,0} =  \tr \int d^3x \Bigg[& D_0 \calz^1 D_0 \bcalz_1 - 2 D\calz^A \bar{D} \bcalz_A - 2 \bar{D} \calz^A D \bcalz_A + \frac{2\pi i }{k} D_0 \calz^A [\bcalz_1, \bcalz_A; \calz^1] \\ \nonumber
    & + \frac{2\pi i}{k} [\calz^1, \calz^A; \bcalz_1] D_0 \bcalz_A - \frac{4\pi^2}{3k^2} [\calz^A, \calz^1; \bcalz_A] [\bcalz_B, \bcalz_1; \calz^B] \\ \nonumber
    & + \frac{16\pi^2}{3k^2} [\calz^A, \calz^1; \bcalz_B] [\bcalz_A, \bcalz_1, \calz^B] + \frac{8 \pi^2}{3k^2} [\calz^A, \calz^B; \bcalz_1] [\bcalz_A, \bcalz_B; \calz^1] \\ \nonumber
    & - \frac{4\pi^2}{3k^2} [\calz^1, \calz^A; \bcalz_1] [\bcalz_B, \bcalz_A; \calz^B] - \frac{4\pi^2}{3k^2} [\calz^B, \calz^A; \bcalz_B] [ \bcalz_1 , \bcalz_A; \calz^1] \\
    & + \frac{ik}{2\pi} \bigg( A_0^L F_{z \bar{z}}^L + A_z^L F_{\bar{z}0}^L + A_{\bar{z}}^L F_{0z}^L + i A_0^L [A_z^L, A_{\bar{z}}^L] - L \to R
    \bigg) \Bigg] \ .
\end{align}
\end{subequations}
However the divergent term $S_{B,-}$ needs to be managed. 
Following the shift in $A_0$ one finds $S_{B,-}$ can be written as 
\begin{equation} \label{eq: divergent pieces}
    S_{B,-} = -   \tr \int d^3x \brac{ 4\bar{D} \calz^1 D \bcalz_1 + 2\partial (\calz^1 \bar{D} \bcalz_1) -2 \bar \partial (\calz^1 D \bcalz_1) } \ .
\end{equation}
The last two terms are a total derivative 
%%%%%%%%%%%%%%%%%%%%%%%%%%%%%%%%%%%%%%%%%%%%%%%
and therefore don't contribute to the dynamics of the theory. However, they will yield a divergent contribution to the energy of any state; we would therefore like a physical origin for the cancellation of these terms.

We can view our field theory as describing the worldvolume dynamics of a stack of M2-branes in a flat orbifolded background. If we turn on a constant 3-form potential
\begin{equation} \label{eq: WZterm}
    \hat{C}_3 = \frac{i}{2\omega^2} dt \wedge d\calz^1_s \wedge d\bcalz_{1,s}
\end{equation}
in the eleven-dimensional spacetime, then using the relation
\begin{equation}
    \calz^1 = \sqrt{\frac{T_{M2}}{2}} \calz^1_s\ ,
\end{equation}
between our ABJM field and eleven-dimensional coordinate the Wess-Zumino term for a single M2-brane in this background is
\begin{align} \nonumber
    S_{WZ} &= T_{M2} \int \hat{C}_3 \\
    &= \frac{2}{\omega^2} \int d^3x \Big( \partial\brac{\calz \bar{\partial} \bcalz} - \bar{\partial}\brac{\calz \partial\bcalz} \Big) \ .
\end{align}
The question of the non-Abelian generalisation of this is slighty subtle. However, if we demand that the only terms that can affect the field theory's dynamics are those that are gauge-invariant under transformations of $\hat{C}_3$\footnote{In other words, for our constant background we should only pick up boundary terms.} then the appropriate generalisation is
\begin{equation}
    \Tilde{S}_{WZ} = \frac{2}{\omega^2} \tr \int d^3x \Big( \partial\brac{\calz \bar{D} \bcalz} - \bar{\partial}\brac{\calz D \bcalz} \Big) \ ,
\end{equation}
which we notice exactly cancels the boundary terms in \eqref{eq: divergent pieces}. Summing the two then leaves us with
\begin{equation}
    \Tilde{S}_{B,-} \equiv S_{B,-} + \Tilde{S}_{WZ} = - 4 \tr \int d^3x \, \bar{D} \calz^1 D \bcalz_1\ .
\end{equation}

%%%%%%%%%%%%%%%%%%%%%%%%%%%%%%%%%%%%%%%%%%%%%

As the integrand is now a squared quantity, we can introduce a complex auxilliary field $H$ in the bifundamental of $U(N)\times U(N)$ and perform a Hubbard-Stratonovich transformation to rewrite the divergent piece as
\begin{equation}
    \inv{\omega^2} \Tilde{S}_{B,-} = \tr \int d^3x \brac{\bar{D} \calz^1 \bar{H} + H D \bcalz_1 + \frac{\omega^2}{4} H \bar{H} } \ ,
\end{equation}
where the equality holds after imposing the equation of motion of the auxilliary field. After taking the scaling limit $\omega\to0$ we reach the fixed-point action
\begin{align} \label{eq: bosonic action} \nonumber
    S_B = \tr \int d^3x \Bigg[& D_0 \calz^1 D_0 \bcalz_1 + H D \bcalz_1 + \bar{D} \calz^1 \bar{H} - 2 D\calz^A \bar{D} \bcalz_A - 2 \bar{D} \calz^A D \bcalz_A \\ \nonumber
    &+ \frac{2\pi i }{k} D_0 \calz^A [\bcalz_1, \bcalz_A; \calz^1] + \frac{2\pi i}{k} [\calz^1, \calz^A; \bcalz_1] D_0 \bcalz_A \\ \nonumber
    &- \frac{4\pi^2}{3k^2} [\calz^A, \calz^1; \bcalz_A] [\bcalz_B, \bcalz_1; \calz^B] + \frac{16\pi^2}{3k^2} [\calz^A, \calz^1; \bcalz_B] [\bcalz_A, \bcalz_1, \calz^B] \\ \nonumber
    &+ \frac{8 \pi^2}{3k^2} [\calz^A, \calz^B; \bcalz_1] [\bcalz_A, \bcalz_B; \calz^1] - \frac{4\pi^2}{3k^2} [\calz^1, \calz^A; \bcalz_1] [\bcalz_B, \bcalz_A; \calz^B] \\ \nonumber
    &- \frac{4\pi^2}{3k^2} [\calz^B, \calz^A; \bcalz_B] [ \bcalz_1 , \bcalz_A; \calz^1] + \frac{ik}{2\pi} \bigg( A_0^L F_{z \bar{z}}^L + A_z^L F_{\bar{z}0}^L + A_{\bar{z}}^L F_{0z}^L \\
    &+ i A_0^L [A_z^L, A_{\bar{z}}^L] - L \to R
    \bigg) \Bigg] \ .
\end{align}
The term quadratic in $H$ has vanished; it becomes a Lagrange multiplier field whose effect is to implement the constraint
\begin{equation} \label{eq: Z constraint}
    \bar{D} \calz^1 = 0 \ ,
\end{equation}
which we recognise as the BPS equation \eqref{eq: holomorphic BPS equation}. Note that for brevity we will write the scalar action as
\begin{align} \nonumber
    S_{\cal{Z}} = \tr \int d^3x \Bigg[& D_0 \calz^1 D_0 \bcalz_1 + H D \bcalz_1 + \bar{D} \calz^1 \bar{H} - 2 D\calz^A \bar{D} \bcalz_A - 2 \bar{D} \calz^A D \bcalz_A \\
    &+ \frac{2\pi i }{k} D_0 \calz^A [\bcalz_1, \bcalz_A; \calz^1] + \frac{2\pi i}{k} [\calz^1, \calz^A; \bcalz_1] D_0 \bcalz_A - V \Bigg] \ .
\end{align}

\subsection{Field Theory Symmetries} \label{sect: symmetries}

As we have emphasised above it is important and interesting to understand the symmetries of the action \eqref{eq: bosonic action}. As the theory at finite $\omega$ is just a rewriting of ABJM, it is guaranteed that any ABJM symmetry without terms that diverge as $\omega\to0$ will either be a symmetry of the fixed-point action or act trivially on it. However, when $\omega=0$ the symmetry structure is much richer.

The spacetime symmetries of the theory can be split into two families. First, motivated by the form of transformations in the original theory, we consider infinitesimal transformations of the form
\begin{subequations}
\begin{align}
    \hat{t} &= t + F(t) \ , \\
    \hat{z} &= z\brac{1 + F'} \ ,
\end{align}
\end{subequations}
in terms of an infinitesimal function $F$ that we leave arbitrary for now. Taking our fields to have the transformations
\begin{subequations}
\begin{align}
    \hat{\calz}^1(\hat{t},\hat{z},\hat{\bar{z}}) &= \brac{1 - \inv{2} F'} \calz^1(t,z,\bar{z}) \ , \\
    \hat{\calz}^A(\hat{t},\hat{z},\hat{\bar{z}}) &= \brac{1 - \inv{2} F'} \calz^A(t,z,\bar{z}) \ , \\
    \hat{H}(\hat{t}, \hat{z}, \hat{\bar{z}}) &= \brac{ \brac{1 - \frac{3}{2}F'}H + 2z F'' D_0 \calz^1 + z F''' \calz^1 }(t,z,\bar{z}) \ , \\
    \hat{A}_0^{L/R} (\hat{t},\hat{z},\hat{\bar{z}}) &= \brac{(1 - F') A_0^{L/R} - z F'' A_z^{L/R} - \bar{z} F'' A_{\bar{z}}^{L/R} } (t,z,\bar{z}) \ , \\
    \hat{A}_z^{L}(\hat{t},\hat{z},\hat{\bar{z}}) &= \brac{ (1 - F') A_z^L + \frac{\pi \bar{z} F''}{k} \calz^1 \bcalz_1  } (t,z,\bar{z}) \ , \\
    \hat{A}_z^{R}(\hat{t},\hat{z},\hat{\bar{z}}) &= \brac{ (1 - F') A_z^R + \frac{\pi \bar{z} F''}{k}  \bcalz_1 \calz^1  } (t,z,\bar{z}) \ , \\
    \hat{A}_{\bar{z}}^{L}(\hat{t},\hat{z},\hat{\bar{z}}) &= \brac{ (1 - F') A_{\bar{z}}^L + \frac{\pi z F''}{k} \calz^1 \bcalz_1  } (t,z,\bar{z}) \ , \\
    \hat{A}_{\bar{z}}^{R}(\hat{t},\hat{z},\hat{\bar{z}}) &= \brac{ (1 - F') A_{\bar{z}}^R + \frac{\pi z F''}{k}  \bcalz_1 \calz^1  } (t,z,\bar{z}) \ ,
\end{align}
\end{subequations}
we find that the change in the action at leading order is
\begin{equation}
    \delta S = -\inv{2} \tr \int d^3x\, F''' \calz^1 \bcalz_1 \ .
\end{equation}
This means that the symmetries of this form are described by the function
\begin{equation} \label{eq: time symmetry}
    F(t) = a + bt + ct^2 \ .
\end{equation}
We see that this is an $SO(2,1)$ subgroup of the original spacetime symmetry group. 
%%%%%%%%%%%%%%%%%%%%
We note that the non-tensorial part of the spatial gauge fields's transformations arises from taking the non-relativistic limit of the standard tensorial relativistic transformations after rewriting our theory in terms of the shifted field \eqref{eq: A0 shift}.
%%%%%%%%%%%%%%%%%%%%

The other family of spacetime transformations that we will work with are the transformations
\begin{subequations} \label{eq: spatial symmetry}
\begin{align}
    \hat{t} &= t \ , \\
    \hat{z} &= z + f(z,t) \ ,
\end{align}
\end{subequations}
for some infinitesimal function $f$. Note that the only requirement we've asked of $f$ is that $\bar{\partial}f=0$. If we take
\begin{subequations} \label{eq: spatial field transformations}
\begin{align}
    \hat{\calz}^1(\hat{t},\hat{z},\hat{\bar{z}}) &= \brac{1 - \pf} \calz^1 (t,z,\bar{z}) \ , \\
    \hat{\calz}^{A} (\hat{t},\hat{z},\hat{\bar{z}}) &= \calz^A (t,z,\bar{z}) \ , \\
    \hat{H}(\hat{t},\hat{z},\hat{\bar{z}}) &= \brac{H + 2 f' D_0 \calz^1 + f'' \calz^1} (t,z,\bar{z}) \ , \\
    \hat{A}_0^{L/R}(\hat{t},\hat{z},\hat{\bar{z}}) &= \brac{A_0^{L/R} - f' A_z^{L/R} - \bar{f}' A_{\bar{z}}^{L/R} } (t,z,\bar{z}) \ , \\
    \hat{A}_z^{L}(\hat{t},\hat{z},\hat{\bar{z}}) &= \brac{ (1 - \pf) A_z^L + \frac{\pi}{k} \bar{f}' \calz^1 \bcalz_1} (t,z,\bar{z}) \ , \\
    \hat{A}_{\bar{z}}^{L}(\hat{t},\hat{z},\hat{\bar{z}}) &= \brac{ (1 - \bpf) A_{\bar{z}}^L + \frac{\pi}{k} f' \calz^1 \bcalz_1} (t,z,\bar{z}) \ , \\
    \hat{A}_z^{R}(\hat{t},\hat{z},\hat{\bar{z}}) &= \brac{ (1 - \pf) A_z^R + \frac{\pi}{k}  \bar{f}' \bcalz_1 \calz^1  } (t,z,\bar{z}) \ , \\
    \hat{A}_{\bar{z}}^{R}(\hat{t},\hat{z},\hat{\bar{z}}) &= \brac{ (1 - \bpf) A_{\bar{z}}^R + \frac{\pi}{k} f'  \bcalz_1 \calz^1 } (t,z,\bar{z}) \ ,
\end{align}
\end{subequations}
we find that the action is invariant. 

There are two interesting symmetry enhancements at play here; the standard spatial symmetry one expects after taking the non-relativistic limit\footnote{That is, spatial translations, Galilean boosts, and rotations.} is enhanced to the two-dimensional Euclidean conformal algebra, and the spatial transformations can be made time-dependent. Spatial transformations with arbitrary time-dependence have been previously studied in massive non-relativistic field theories coupled to background gauge fields \cite{Son:2013rqa, Jensen:2014aia}. However, upon adding a Chern-Simons term for the gauge field these symmetries are lost unless the massless limit of matter in the theory is taken; this reproduces the non-dynamical theory discussed in section \ref{sect: toy}. The novelty here is that introduction of a matter field with two temporal derivative kinetic term allows us to retain this structure in a dynamical field theory. It is also this derivative structure that allows for the infinite-dimensional extension of the spatial symmetry.

It will be convenient to disentangle the spatial and temporal symmetries. If we combine the two types of transformations and take
\begin{equation}
    f = - z F'
\end{equation}
then we find a purely temporal symmetry
\begin{subequations} \label{eq: temporal symmetry}
\begin{align}
    \hat{t} &= t + F(t) \ , \\
    \hat{z} &= z \ ,
\end{align}
\end{subequations}
(with $F$ as in \eqref{eq: time symmetry}) under which the fields have the transformations
\begin{subequations}
\begin{align}
    \hat{\calz}^1(\hat{t},\hat{z},\hat{\bar{z}}) &= \brac{1 + \inv{2}F'} \calz^1(t,z,\bar{z}) \ , \\
    \hat{\calz}^{A}(\hat{t},\hat{z},\hat{\bar{z}}) &= \brac{1 - \inv{2}F'} \calz^A(t,z,\bar{z}) \ , \\
    \hat{H}(\hat{t},\hat{z},\hat{\bar{z}}) &= \brac{1 - \frac{3}{2}F'} H(t,z,\bar{z}) \ , \\
    \hat{A}^{L/R}_0(\hat{t},\hat{z},\hat{\bar{z}}) &= (1-F') A_0^{L/R}(t,z,\bar{z}) \ , \\
    \hat{A}_z^{L/R} (\hat{t},\hat{z},\hat{\bar{z}}) &= A_z^{L/R}(t,z,\bar{z}) \ , \\
    \hat{A}_{\bar{z}}^{L/R} (\hat{t},\hat{z},\hat{\bar{z}}) &= A_{\bar{z}}^{L/R}(t,z,\bar{z}) \ .
\end{align}
\end{subequations}

In addition to the spacetime symmetries, we have the R-symmetries
\begin{subequations} \label{eq: U(1) sym}
\begin{align}
    \hat{\calz}^1 &= e^{i\alpha} \calz^1 \ , \\
    \hat{H} &= e^{i\alpha} H \ , \\
    \hat{\calz}^A &= \calz^A \ ,
\end{align}
\end{subequations}
and
\begin{subequations} \label{eq: SU(3) sym}
\begin{align}
    \hat{\calz}^1 &= \calz^1 \ , \\
    \hat{H} &=  H \ , \\
    \hat{\calz}^A &= \tensor{\cal{R}}{^A_B} \calz^B \ ,
\end{align}
\end{subequations}
for $\cal{R}\in SU(3)$. We also retain the global $U(1)_b\subset U(N) \times U(N)$ baryon number symmetry
\begin{subequations}
\begin{align}
    \hat{\calz}^1 &= e^{i\beta} \calz^1 \ , \\
    \hat{H} &= e^{i\beta} H \ , \\
    \hat{\calz}^A &= e^{i \beta} \calz^A \ .
\end{align}
\end{subequations}
The R-symmetries form a $U(1)_R\times SU(3)$ subgroup of the original $SU(4)$ R-symmetry of ABJM\footnote{Note that a $U(1)_b$ transformation needs to be simultaneously performed with the transformation \eqref{eq: U(1) sym} for the latter to be contained within the original R-symmetry group.}. We note that all the symmetries found in this section can be extended to the full supersymmetric action, with the Fermionic transformations detailed in appendix \ref{sect: fermion symmetries}.

\subsection{Conserved Currents} \label{sect: currents}

We can find the conserved currents associated with the transformations using the standard Noether procedure of promoting the transformation parameters to arbitrary functions. Let us first do this for the temporal symmetry \eqref{eq: temporal symmetry}. It will be convenient to work with the modified transformations
\begin{subequations}
\begin{align}
    \hat{A}_z^{L/R}(\hat{t},\hat{z},\hat{\bar{z}}) &= \brac{A^{L/R}_z - \partial F A^{L/R}_0}(t,z,\bar{z})  \ , \\
    \hat{A}_{\bar{z}}^{L/R}(\hat{t},\hat{z},\hat{\bar{z}}) &= \brac{A^{L/R}_{\bar{z}} - \bar{\partial} F A^{L/R}_0}(t,z,\bar{z}) \ ,
\end{align}
\end{subequations}
after introducing spatial dependence into $F$ in order to maintain gauge-invariance throughout the calculation. The conserved currents are then
\begin{subequations}
\begin{align}
    j^0_{(a)} &= \tr\brac{D_0 \calz^1 D_0 \bcalz_1 + 2 D\calz^A \bar{D} \bcalz_A + 2 \bar{D} \calz^A D \bcalz_A + V } \ , \\
    j_{(a)} &= \tr\brac{H D_0 \bcalz_1 - 2 D_0 \calz^A \bar{D} \bcalz_A - 2 \bar{D} \calz^A D_0 \bcalz_A }\ , \\
    \bar{j}_{(a)} &= \tr\brac{ D_0 \calz^1 \bar{H} - 2 D\calz^A D_0 \bcalz_A - 2 D_0 \calz^A D \bcalz_A } \ ,
\end{align}  
\end{subequations}
for the transformation $F = a$,
\begin{subequations}
\begin{align}
    j^0_{(b)} &= \tr\brac{t \bigg( D_0 \calz^1 D_0 \bcalz_1 + 2 \brac{D \calz^A \bar{D} \bcalz_A + \bar{D} \calz^A D \bcalz_A} + V \bigg) - \inv{2} \brac{\calz^1 D_0 \bcalz_1 + D_0 \calz^1 \bcalz_1} } \ , \\
    j_{(b)} &= \tr \brac{ t \bigg( H D_0 \bcalz_1 - 2 D_0 \calz^A \bar{D} \bcalz_A - 2 \bar{D} \calz^A D_0 \bcalz_A \bigg) - \calz^A \bar{D} \bcalz_A - \bar{D} \calz^A \bcalz_A - \inv{2} H \bcalz_1 } \ , \\
    \bar{j}_{(b)} &= \tr \brac{ t \bigg( D_0 \calz^1 \bar{H} - 2 D_0 \calz^A D \bcalz_A - 2 D \calz^A D_0 \bcalz_A \bigg) - \calz^A D \bcalz_A - D \calz^A \bcalz_A - \inv{2} \calz^1 \bar{H} } \ ,
\end{align}
\end{subequations}
for the transformation $F = bt$, and
\begin{subequations}
\begin{align}
    j^0_{(c)} &= \tr\brac{t^2 \bigg( D_0 \calz^1 D_0 \bcalz_1 + 2 \brac{D \calz^A \bar{D} \bcalz_A + \bar{D} \calz^A D \bcalz_A} + V \bigg) - t D_0 \brac{\calz^1 \bcalz_1} + \calz^1 \bcalz_1 } \ , \\
    j_{(c)} &= \tr \brac{ t^2 \bigg( H D_0 \bcalz_1 - 2 D_0 \calz^A \bar{D} \bcalz_A - 2 \bar{D} \calz^A D_0 \bcalz_A \bigg) - t \brac{2 \calz^A \bar{D} \bcalz_A +2 \bar{D} \calz^A \bcalz_A + H \bcalz_1} } \ , \\
    \bar{j}_{(c)} &= \tr \brac{ t^2 \bigg( D_0 \calz^1 \bar{H} - 2 D_0 \calz^A D \bcalz_A - 2 D \calz^A D_0 \bcalz_A \bigg) - t \brac{2 \calz^A D \bcalz_A +2 D \calz^A \bcalz_A + \calz^1 \bar{H} } } \ ,
\end{align}
\end{subequations}
for the transformation $F = ct^2$. Note that we have used the equations of motion to write the currents as above.

We can perform the same calculation for the spatial symmetry \eqref{eq: spatial symmetry}, where we similarly modify the gauge field transformations to
\begin{subequations}
\begin{align}
    \hat{A}_z^{L}(\hat{t},\hat{z},\hat{\bar{z}}) &= \brac{(1 - \partial f) A_z^L - \partial \bar{f} A_{\bar{z}}^L + \frac{\pi \bar{f}'}{k} \calz^1 \bcalz_1}(t,z,\bar{z}) \ , \\
    \hat{A}_{\bar{z}}^{L}(\hat{t},\hat{z},\hat{\bar{z}}) &= \brac{(1 - \bar{\partial} \bar{f}) A_{\bar{z}}^L - \bar{\partial} f A_{z}^L + \frac{\pi f'}{k} \calz^1 \bcalz_1}(t,z,\bar{z}) \ ,
\end{align}
\end{subequations}
and analogously for $(A_z^R,A_{\bar{z}}^R)$. As the action is invariant for any time-dependent holomorphic function $f(z,t)$, the only non-vanishing terms after taking $f$ to have arbitrary coordinate dependence take the form
\begin{equation}
    \delta S = - \int d^3x \brac{ \bar{\partial} f\, T + \partial \bar{f} \, \bar{T} 
    } \ ,
\end{equation}
so the associated conservation laws are
\begin{equation} \label{eq: holomorphic constraint}
    0 = \bar{\partial} T
\end{equation}
and its complex conjugate. A brief computation gives
\begin{equation}
    T = \tr \brac{\calz^1 D \bar{H} + 4 D\calz^A D \bcalz_A} \ .
\end{equation}
Since the symmetry can be taken to have arbitrary time dependence the current has no temporal component, so there is no codimension-1 conserved charge associated with the symmetry. We should interpret the conservation law \eqref{eq: holomorphic constraint} as a constraint we must impose on the physical states of the theory. At first glance, it appears somewhat strange to have gone from a theory possessing codimension-1 charges associated with spatial translations and rotations to one where these transformations are still symmetries of the theory but the analogous charges vanish. To understand this better it is instructive to directly perform the limit for the spatial momentum in ABJM,
\begin{equation}
    \hat{P}_i = \tr \int_{\hat{\Sigma}} d^2 \hat{x} \brac{D_0 \hat{\calz}^M D_i \hat{\bcalz}_M + D_i \hat{\calz}^M D_0 \hat{\bcalz}_M} \ .
\end{equation}
Shifting the temporal component of the gauge field as in \eqref{eq: A0 shift} and performing the scaling \eqref{eq: scaling 2} gives
\begin{subequations}
\begin{align}
    \hat{P}_i &= \inv{\omega} P_i \ , \\ \nonumber
    P_i &= \tr \int_{\Sigma} d^2 x \bigg[D_0 \calz^1 D_i \bcalz_1 + D_i \calz^1 D_0 \bcalz_1 + \frac{2\pi i}{k} \bigg([\calz^1 , \calz^A ; \bcalz_1] D_i \bcalz_A \\
    & \hspace{1.0cm} + D_i \calz^A [ \bcalz_1 , \bcalz_A ; \calz^1]\bigg) \bigg] + O(\omega^2) \ .
\end{align}
\end{subequations}
Note that since we have scaled the spatial coordinates as $\hat{x}^i = \omega x^i$ it is natural for the momentum to have the overall scaling above. Taking the limit $\omega \to 0$ in $P_i$, we can use the constraint \eqref{eq: Z constraint} to write the $z$-component of the momentum as
\begin{equation}
    P_z = \tr \int_{\Sigma} d^2 x \bigg[ D \calz^1 D_0 \bcalz_1 + \frac{2\pi i}{k} \bigg( [\calz^1, \calz^A ; \bcalz_1] D\bcalz_A + D\calz^A [\bcalz_1, \bcalz_A ; \calz^1] \bigg) \bigg] \ ,
\end{equation}
which after integrating by parts and again using the constraint is
\begin{align} \nonumber
    P_z = \tr \int_{\Sigma} d^2x \bigg[& \partial\brac{\calz^1 D_0 \bcalz_1} + i \calz^1 \bcalz_1 F^L_{0z} - i \bcalz_1 \calz^1 F^R_{0z} \\
    &+ \frac{2\pi i}{k} \bigg( [\calz^1, \calz^A ; \bcalz_1] D\bcalz_A + D\calz^A [\bcalz_1, \bcalz_A ; \calz^1] \bigg) 
    \bigg] \ .
\end{align}
However, when the equations of motion \eqref{eq: equations of motion} are satisfied\footnote{We set the Fermionic terms to zero here.} the final four terms cancel, leaving us with
\begin{equation}
    P_z = i \tr \int_{\partial \Sigma} d\bar{z} \, \calz^1 D_0 \bcalz_1 \ .
\end{equation}
The limit has taken the codimension-1 charge to a codimension-2 charge evaluated at spatial infinity. We could do the same for the spatial rotation
\begin{equation}
    \hat{Q}_{ij} = \tr \int_{\hat{\Sigma}} d^2 \hat{x} \bigg[ \hat{x}^i \brac{D_0 \hat{\calz}^M D_j \hat{\bcalz}_M + D_j \hat{\calz}^M D_0 \hat{\bcalz}_M} - \hat{x}^j \brac{D_0 \hat{\calz}^M D_i \hat{\bcalz}_M + D_i \hat{\calz}^M D_0 \hat{\bcalz}_M} \bigg] \ .
\end{equation}
Following the same steps as above, we find the conserved charge
\begin{align}
    Q_{z\bar{z}} = -\tr \int_{\partial\Sigma} \brac{ d\bar{z} \, z \calz^1 D_0 \bcalz_1 + dz \, \bar{z} D_0 \calz^1 \bcalz_1} - i\tr \int_{\Sigma} d^2 x \brac{\calz^1 D_0 \bcalz_1 - D_0 \calz^1 \bcalz_1} \ .
\end{align}
It appears that in this case we've found a non-trivial codimension-1 conserved charge. However, as we'll see momentarily this is just the conserved charge associated with the R-symmetry transformation \eqref{eq: U(1) sym}, reflecting the fact that the transformations of $\calz^1$ under the ABJM spatial rotation after the $\omega\to0$ scaling and under the rotation in the family of transformations \eqref{eq: spatial field transformations} differ by the action of this global symmetry. More generally, applying Noether's theorem for a time-independent transformation of the form \eqref{eq: spatial symmetry} gives the charge
\begin{equation}
    Q[f]  = i \tr \int_{\partial\Sigma} \brac{d \bar{z} \, f \calz^1 D_0 \bcalz_1 - dz \, \bar{f} D_0 \calz^1 \bcalz_1} \ .
\end{equation}
Since these take the form of boundary terms they cannot generate a symmetry transformation on the bulk phase space of the theory. It therefore seems natural to interpret these transformations as gauge redundancies.

%%%%%%%%%%%%%%%%%%%%%%%%%%%%%%%%%%%%%%%%%%%%%%%%
The relativistic spacetime symmetry has been broken into a 'physical' $SO(2,1)$ factor with non-trivial conserved charges and the 'unphysical' spatial symmetries, which have no conserved charges. The physical interpretation of this is that the non-relativistic limit enforces a constraint on the theory, with the dynamics reducing to quantum mechanics on the constraint surface. The $SO(2,1)$ factor, which is the one-dimensional global conformal group, can then be interpreted as the spacetime symmetry of the quantum mechanical system. The spatial symmetries become symmetries of the constraint equations, which from the quantum mechanics perspective are internal symmetries: the fact that they can be made time-dependent can then be interpreted as a gauging of the symmetries.
%%%%%%%%%%%%%%%%%%%%%%%%%%%%%%%%%%%%%%%%%%%%%%%%

Finally, we give the R-symmetry and baryon number currents. The $U(1)_R$ R-symmetry transformation \eqref{eq: U(1) sym} has the current
\begin{subequations}
\begin{align}
    j^0 &= i \tr\brac{\calz^1 D_0 \bcalz_1 - D_0 \calz^1 \bcalz_1} \ , \\
    j &= -i \tr \brac{H \bcalz_1} \ , \\
    \bar{j} &= i \tr \brac{\calz^1 \bar{H}} \ ,
\end{align}
\end{subequations}
and the $SU(3)$ symmetry \eqref{eq: SU(3) sym} has
\begin{subequations}
\begin{align}
    \tensor{\brac{J^0}}{^A_B} &= \frac{2\pi}{k} \tr \brac{[\calz^1 , \calz^A ; \bcalz_1] \bcalz_B - \calz^A [ \bcalz_1, \bcalz_B ; \calz^1] - \frac{2}{3}\delta^A_B [ \calz^1 , \calz^C ; \bcalz_1] \bcalz_C} \ , \\
    \tensor{J}{^A_B} &= 2i \tr \brac{ \bar{D} \calz^A \bcalz_B - \calz^A \bar{D} \bcalz_B - \inv{3} \delta^A_B \brac{\bar{D} \calz^C \bcalz_C - \calz^C \bar{D}\bcalz_C} } \ , \\
    \tensor{\bar{J}}{^A_B} &= 2i \tr \brac{ D \calz^A \bcalz_B - \calz^A D \bcalz_B - \inv{3} \delta^A_B \brac{D \calz^C \bcalz_C - \calz^C D\bcalz_C} } \ .
\end{align}
\end{subequations}
The $U(1)_b$ baryon number current is
\begin{subequations}
\begin{align}
    j_b^0 &= i \tr\brac{\calz^1 D_0 \bcalz_1 - D_0 \calz^1 \bcalz_1 - \frac{2\pi i}{k}  \brac{[\calz^1 , \calz^A ; \bcalz_1] \bcalz_B - \calz^A [ \bcalz_1, \bcalz_B ; \calz^1] } } \ , \\
    j_b &= i \tr \brac{ 2\bar{D} \calz^A \bcalz_B - 2\calz^A \bar{D} \bcalz_B - i H \bcalz_1 } \ , \\
    \bar{j}_b &= i \tr \brac{ 2D \calz^A \bcalz_B - 2 \calz^A D \bcalz_B + \calz^1 \bar{H}} \ .
\end{align}
\end{subequations}

%%%%%%%%%%%%%%%%%%%%%%%%%%%%%%%%%%%%%%%%%%
\subsection{A Limit on \texorpdfstring{$\bb{R}\times S^2$}{R x S2}}
While we have focused on ABJM on a flat background above, we may also wonder if a similar limit can be taken when the theory is defined on $\bb{R}\times S^2$. The main difference here is the presence of the conformal coupling for the scalar fields, which modifies the relativistic scalar action to
\begin{equation}
    \hat{S}_{\calz} = - \tr\int d^3x \sqrt{-g} \brac{D_{\mu} \hcalz^M D^{\mu} \hbcalz_M + \frac{\cal{R}[g]}{8} \hcalz^M \hbcalz_M + \hat{V}} \ ,
\end{equation}
where $\cal{R}[g]$ is the Ricci scalar of $g$. We will take our metric to be
\begin{align}
    ds^2 &=  -dt^2  + \omega^2 \Omega^{2} dx^i dx^i \ , \\
    \Omega &= \frac{2 R}{1 + x^i x^i} \ ,
\end{align}
where as before we can use a constant dilatation to arrange the powers of $\omega$ in this way. Note that here we have taken the non-relativistic limit by shrinking the radius of the sphere, therefore retaining its form. An alternative approach would be to take the limit without changing the radius by zooming in on a point; this would then flatten out the sphere and give the limit considered previously. 

The derivative terms on this background are
\begin{equation}
    \hat{S}_{\calz,\partial} = \tr\int d^3x  \brac{
    \omega^2 \Omega^2 D_0 \hcalz^M D_0 \hbcalz_M -  D_i \hcalz^M D_i \hbcalz_M - \frac{\Omega^2}{4 R^2} \hcalz^M \hbcalz_M
    } \ ,
\end{equation}
with the conformal coupling acting as a mass term for the scalars. The calculation for $\calz^A$ proceeds as before: taking $\hcalz^A = \calz^A$ allows the $\omega \to 0$ limit to be taken for these fields, yielding
\begin{equation}
    S_A = -\tr\int d^3x \brac{ D_i \calz^A D_i \bcalz_A + \frac{\Omega^2}{4R^2} \calz^A \bcalz_A} \ .
\end{equation}
However, if we try and take the same scaling as before for $\calz^1$ we find that the mass term is divergent. Taking the $\omega\to0$ limit would then impose $\calz^1=0$ and the theory would be dynamically trivial. If we instead take our reparameterisation to be
\begin{equation} \label{eq: Z1 sphere reparam}
    \hcalz^1 = \omega^{-1/2} e^{- \frac{i t}{2\omega R}} \calz^1\ ,
\end{equation}
then we find
\begin{equation}
    \hat{S}_1 = \tr\int d^3x \brac{ \Omega^2 \omega D_0 \calz^1 D_0 \bcalz_1 + \frac{i \Omega^2}{2 R} \brac{D_0 \calz^1 \bcalz_1 - \calz^1 D_0 \bcalz_1} - \omega^{-1} D_i \calz^1 D_i \bcalz_1 } ;
\end{equation}
the only divergent term is now the same as the flat case, and the same arguments (including the shifts in $A_0^L$ and $A_0^R$) give the derivative terms
\begin{equation}
    S_1 = \tr\int d^3x \brac{\frac{i \Omega^2}{2 R} \brac{D_0 \calz^1 \bcalz_1 - \calz^1 D_0 \bcalz_1} + H D \bcalz_1 + \bar{D} \calz^1 \bar{H} }\ ,
\end{equation}
in the $\omega \to 0$ limit, after regulating the $O(\omega^{-1})$ divergence. The form \eqref{eq: Z1 sphere reparam} of the reparameterisation could have been anticipated from the $\inv{2}$-BPS solution of ABJM on $\bb{R}\times S^2$ \cite{Ezhuthachan:2011kf}, which features the same oscillatory term. 

From the scalings of the fields, we see that the only terms from the relativistic theory's potential that survive the $\omega\to0$ limit are those quartic in $\calz^1$. These are finite, so in the $\omega\to0$ limit we have
\begin{equation} \label{eq: sphere potential}
    S_V = - \frac{4\pi^2 }{k^2}\tr \int d^3x \, \Omega^2 [\calz^1 , \calz^A ; \bcalz_1] [\bcalz_1 , \bcalz_A ; \calz^1] \ .
\end{equation}
As the shifts in the timelike components of the gauge fields are
\begin{subequations}
\begin{align}
    \delta A_0^L &= - \frac{2\pi}{\omega k} \calz^1 \bcalz_1 \ , \\
    \delta A_0^R &= - \frac{2\pi}{\omega k} \bcalz_1 \calz^1 \ , 
\end{align}
\end{subequations}
we pick up the terms
\begin{align} \nonumber
    S'_A = \tr \int d^3x \, \Omega^2 \bigg(& \frac{4\pi^2}{k^2} [\calz^1 , \calz^A ; \bcalz_1] [\bcalz_1 , \bcalz_A ; \calz^1] - \frac{2\pi \omega}{k} \Big([\calz^1, \calz^A ; \bcalz_1] D_0 \bcalz_A \\ 
    &+ D_0 \calz^A [\bcalz_1, \bcalz_A ; \calz^1]\Big) \bigg)
\end{align}
from making the shift in the kinetic term of $\calz^A$. In the $\omega\to0$ limit only the first term remains, and we see that it exactly cancels the potential \eqref{eq: sphere potential}. The scalar action is therefore
\begin{align} \nonumber
    S_{\calz} = \tr \int d^3x \bigg(&
    \frac{i \Omega^2}{2 R} \brac{D_0 \calz^1 \bcalz_1 - \calz^1 D_0 \bcalz_1} + H D \bcalz_1 + \bar{D} \calz^1 \bar{H} - D \calz^A \bar{D} \bcalz_A \\
    &- \bar{D} \calz^A D \bcalz_A - \frac{\Omega^2}{4R^2} \calz^A \bcalz_A \bigg) \ .
\end{align}
Unlike the flat-space limit this is quadratic in the scalar fields. It is easy to see that the Gauss law and Lagrange multiplier constraints still leads to a Hitchin-like system:
\begin{subequations}
\begin{align}
    \bar D \calz^1 & =0\\
    F^L_{z\bar z} & = \frac{i\pi \Omega^2}{kR} \calz^1\bcalz^1 \\
    F^R_{z\bar z} & = \frac{i\pi\Omega^2}{kR} \bcalz^1\calz^1\ .
\end{align}  
\end{subequations}
Note that there is an explicit dependence on $z = x^1+ix^2$ as $\Omega=2R/(1+z\bar z)$. As the relativistic theories on $\bb{R}^{1,2}$ and $\bb{R}\times S^2$ are related by the state-operator map, it would be interesting to see if something similar applies here or if the theories are wholly unrelated. In addition it is not clear if it still has an infinite-dimensional set of symmetries. We leave these issues  to future work.

%%%%%%%%%%%%%%%%%%%%%%%%%%%%%%%%%%%%%%%%%%

\section{Eleven-Dimensional Membrane Newton-Cartan Gravity and the M2-Brane} \label{sect: 11d gravity}

\subsection{A Brief Review of Non-Relativistic Eleven-Dimensional Supergravity}

We have constructed our field theory by taking a non-relativistic limit of ABJM. Since ABJM is dual to M-theory on an asymptotically $AdS_4\times S^7/\bb{Z}_k$ background, we would like to try and understand whether a similar duality holds between the non-relativistic field theory and a non-relativistic limit of eleven-dimensional supergravity. A limit of the Bosonic sector of eleven-dimensional supergravity that naturally couples to membranes, known as eleven-dimensional membrane-Newton-Cartan geometry, was found in \cite{Blair:2021waq}; we will review the key features of their limit before applying it to the M2-brane metric in the next section. For a general overview of recent progress in understanding non-relativistic gravity see \cite{Hansen:2020pqs, Hartong:2022lsy}.

The Bosonic field content of eleven-dimensional supergravity is the metric $\hat{g}_{\mn}$ and a 3-form field $\hat{C}_3$. In order to define a non-relativistic limit of eleven-dimensional supergravity, we choose a vielbein  for our spacetime which we partition into two sets $\{ \hat{E}^a, \hat{E}^I\}$, with $a\in\{0,1,2\}$ and $I\in\{3,4,...,10\}$. We can then introduce a dimensionless parameter $c$ and redefine our veilbein as
\begin{subequations}
\begin{align}
    \hat{E}^a &= c \hat{\tau}^a \ , \\
    \hat{E}^I &= \inv{\sqrt{c}} \hat{e}^I \ .
\end{align}
\end{subequations}
When working with a specific metric we will implement such a transformation by choosing a set of coordinates and introducing the parameter $c$ using a combination of coordinate and parameter scalings. This means that the components of our vielbein will generically have $c$-dependence; we'll assume that we've scaled quantities such that $\hat{\tau}^a$ and $\hat{e}^I$ admit a well-defined Taylor series in $\inv{c^3}$. As will be seen later, this is the case for the non-relativistic limit of the M2-brane metric. The corresponding orthonormal frame $\{E_a, E_I \}$ must scale as
\begin{subequations}
\begin{align}
    E_a &= \inv{c} \hat{\tau}_a \ , \\
    E_I &= \sqrt{c} \hat{e}_I \ ,
\end{align}
\end{subequations}
so that bases remain dual to each other. In terms of the metric, this means that we have
\begin{subequations}
\begin{align} \label{eq: metric split}
    \hat{g}_{\mn} &= c^2 \hat{\tau}_{\mn} + \inv{c} \hat{H}_{\mn} \ , \\
    \hat{g}^{\mn} &= c \hat{H}^{\mn} + \inv{c^2} \hat{\tau}^{\mn}\ ,
\end{align}
\end{subequations}
where we've defined
\begin{subequations}
\begin{align}
    \hat{\tau}_{\mn} &= \eta_{ab} \hat{\tau}^a_{\mu} \hat{\tau}^b_{\nu} \ , \\
    \hat{\tau}^{\mn} &= \eta^{ab} \hat{\tau}_a^{\mu} \hat{\tau}_b^{\nu} \ , \\
    \hat{H}_{\mn} &= \delta_{IJ} \hat{e}^I_{\mu} \hat{e}^J_{\nu} \ , \\
    \hat{H}^{\mn} &= \delta^{IJ} \hat{e}_I^{\mu} \hat{e}_J^{\nu} \ .
\end{align}
\end{subequations}
We can now expand all quantities in $\inv{c^3}$ using the notation
\begin{subequations} \label{eq: hatted variables}
\begin{align}
    \hat{\tau}^a &= \tau^a + \inv{c^3} m^a + O\brac{\inv{c^6}} \ , \\
    \hat{\tau}_a &= \tau_a + \inv{c^3} M_a +  O\brac{\inv{c^6}} \ , \\
    \hat{e}^I &= e^I + \inv{c^3} \pi^I +  O\brac{\inv{c^6}} \ , \\
    \hat{e}_I &= e_I + \inv{c^3} \Phi_I + O\brac{{\inv{c^6}}} \ ,
\end{align}
\end{subequations}
so the metric and its inverse take the form
\begin{subequations}
\begin{align}
    g_{\mn} &= c^2 \tau_{\mn} + \inv{c} \brac{H_{\mn} + 2 \eta_{ab} \tau^a_{(\mu} m_{\nu)}^b } + O\brac{ \inv{c^4}} \ , \\
    g^{\mn} &= c H^{\mn} + \inv{c^2} \brac{\tau^{\mn} + 2\delta^{IJ} e_{I}^{(\mu} \Phi^{\nu)}_J } + O\brac{\inv{c^{5}}} \ .
\end{align}
\end{subequations}

At finite $c$ the local symmetry algebra of the vielbein is $\frak{so}(1,10)$, with factors of $c$ inserted to match the expansion in the metric. However, as we will ultimately be interested in taking the $c\to \infty$ limit we will only be interested transformations of our variables that are independent of $c$. While it's clear that the $\frak{so}(1,2) \oplus \frak{so}(8)$ subalgebra consisting of Lorentz transformations of $\tau^a$ and rotations of $e^I$ satisfies this condition, it is less obvious which of the transformations that mix the two are retained in the limit. Expanding the invertibility condition $\delta^{\mu}_{\nu} = g^{\mu\rho} g_{\rho\nu}$ in powers of $c$ gives the relations
\begin{subequations}
\begin{align}
    \tau^a(e_I) &= 0 \ , \\
    e^I(\tau_a) &= 0 \ , \\
    \tau^a(\tau_b) &= \delta^a_b \ , \\
    e^I(e_J) &= \delta^I_J \ , \\
    \delta &= \tau_a \otimes \tau^a + e_I \otimes e^I \ , \\
    M_a &= - \brac{m^b(\tau_a) \tau_b + \pi^I(\tau_a) e_I} \ , \\
    \Phi_I &= - \brac{\pi^J (e_I) e_J + m^a(e_I) \tau_a} \ .
\end{align}
\end{subequations}
It's easy to see that these are invariant under the reparameterisations
\begin{subequations} \label{eq: g boosts 1}
\begin{align}
    e^{\prime I} &= e^I + \lambda^I_a \tau^a \ , \\
    \tau_a' &= \tau_a - \lambda^I_a e_I \ ,
\end{align}
\end{subequations}
which we recognise as the leading-order term in the non-relativistic expansion of a local Lorentz boost. These act non-trivially on $H_{\mn}$ and $\tau^{\mn}$, changing them to
\begin{subequations} \label{eq: local galilean boosts}
\begin{align}
    H'_{\mn} &= H_{\mn} + 2 \delta_{IJ} \lambda_a^I e^J_{(\mu} \tau^a_{\nu)} + \delta_{IJ} \lambda_a^I \lambda_b^J \tau_{\mu}^a \tau_{\nu}^b \ , \\
    \tau^{\prime \mn} &= \tau^{\mn} - 2\eta^{ab} \lambda^I_a \tau_b^{(\mu} e_I^{\nu)} + \eta^{ab} \lambda^I_a \lambda^J_b e^{\mu}_I e^{\nu}_J \ .
\end{align}
\end{subequations}
However, also taking the subleading terms to have the transformations
\begin{subequations} \label{eq: g boosts 2}
\begin{align}
    m^{\prime a} &= m^a - \lambda^a_I e^I - \inv{2} \lambda^a_I \lambda^I_b\tau^b \ , \\
    \Phi^{\prime}_I &= \Phi_I + \lambda^a_I v_a - \inv{2} \lambda^a_I \lambda^J_a e_J \ ,
\end{align}
\end{subequations}
we see that both $g_{\mn}$ and $g^{\mn}$ are invariant at subleading order in $c$. The transformations \eqref{eq: g boosts 1} and \eqref{eq: g boosts 2} form a local invariance of the system known as a local Galilean boost. The boosts have the unusual property that they alter quantities with all local indices contracted. This is a manifestation of the fact that both $H_{\mn}$ and $\tau^{\mn}$ are sections of quotients of tensor bundles \cite{Bergshoeff:2023rkk} and are therefore only well-defined up to an equivalence class. The variables $\tau_{\mn}$ and $H^{\mn}$ are, however, sections of genuine tensor bundles and therefore don't transform under the local boosts.

To find the dynamics of the theory in the non-relativistic limit we must specify a prescription for the expansion of the 3-form field $\hat{C}_3$ after the introduction of $c$. This turns out to be highly constrained by the requirement that divergent terms in the action should cancel on physical solutions, and forces us to take decompose $\hat{C}_3$ as
\begin{equation} \label{eq: C3 decomp}
    \hat{C}_3 = - \frac{c^3}{6} \epsilon_{abc} \tau^a \wedge \tau^b \wedge \tau^c + C_3 + \inv{c^3} \tilde{C}_3 + O\brac{\inv{c^6}}\ .
\end{equation}
With this, the action takes the schematic form
\begin{equation}
    \hat{S}_{11d} = c^3 S_3 + S_0 + O\brac{\inv{c^3}} \ ;
\end{equation}
requiring that the divergent piece cancels gives us the constraint
\begin{subequations}
\begin{align} \label{eq: constraint}
    \Omega H^{\mu_1 \nu_1} ... H^{\mu_4 \nu_4} F_{\nu_1 ... \nu_4} &= - \inv{144} \epsilon^{\mu_1 ... \mu_{11}} \epsilon_{abc} F_{\mu_5 ... \mu_9} \tau_{\mu_9}^a \tau_{\mu_{10}}^b \tau_{\mu_{11}}^c \ ,
\end{align}
where $F_4$ is the field strength of $C_3$ and we define $\Omega$ by
\begin{align}
    \Omega^2 &= - \inv{3! 8!} \epsilon^{\mu_1 ... \mu_{11}} \epsilon^{\nu_1 ... \nu_{11}} \tau_{\mu_1 \nu_1} ... \tau_{\mu_3 \nu_3} H_{\mu_4 \nu_4} ... H_{\mu_{11} \nu_{11}} \ . 
\end{align}
\end{subequations}
The constraint is imposed dynamically using a totally antisymmetric Hubbard-Stratonovich field\footnote{We thank Chris Blair for discussions on this point.} $G_{IJKL}$ \cite{Bergshoeff:2023igy}. The action is then finite and the $c\to\infty$ limit can be taken. In this limit the subleading fields are absorbed by $G_{IJKL}$ and drop out of the dynamics (see appendix \ref{sect: 11d limit} for further details). This means that the transformations \eqref{eq: g boosts 1}, and hence also the corresponding metric transformations, become local invariances of the theory, as can be verified from the action upon also taking the 3-form $C_3$ and the Hubbard-Stratonovich field $G_{IJKL}$ to have the infinitesimal transformations
\begin{subequations}
\begin{align}
    \delta C_{\mu\nu\rho} &= - 3 \epsilon_{abc} \lambda_I^a e^{I}_{[\mu} \tau^b_{\nu} \tau^{c}_{\rho]} \ , \\
    \delta G_{IJKL} &= - 4 \lambda^a_{[I} e^{\mu}_{J} e^{\nu}_K e^{\rho}_{L]} \tau^{\sigma}_a F_{\mu \nu\rho\sigma} \ .
\end{align}
\end{subequations}

There is some freedom in how $c$ is defined since it will not be present in the theory after the limit is taken. For instance, redefining $c$ by an overall scale
\begin{equation}
    c = \lambda c_{\lambda}
\end{equation}
and taking $c_{\lambda}\to\infty$ will yield the same limit as taking $c\to\infty$ for any $\lambda>0$, since the equations of motion of the system arise from the $c$-independent piece of the action. This corresponds to an emergent dilatation symmetry that takes
\begin{subequations}
\begin{align}
    \tau_{\mn} &\to \lambda^2 \tau_{\mn} \  , \\
    H^{\mn} &\to \lambda H^{\mn} \ .
\end{align}
\end{subequations}
Since $C_3$ appears with no overall factor of $c$ it should not transform under the dilatation, but we must take $G_{IJKL}$ to have the transformation
\begin{equation}
    G_{IJKL} \to \lambda^{-3} G_{IJKL} \ .
\end{equation}
In fact, we can go further than this; the action is invariant under the infinitesimal transformations
\begin{subequations}
\begin{align}
    \delta \tau_{\mn} &= 2 \Lambda(x) \tau_{\mn} \ , \\
    \delta H^{\mn} &= \Lambda(x) H^{\mn} \ , \\
    \delta G_{IJKL} &= - 3 \Lambda(x) G_{IJKL} \ ,
\end{align}
\end{subequations}
for any function $\Lambda(x)$, so the obvious dilatations are enhanced to a local symmetry of the non-Lorentzian theory.

\subsection{Non-Relativistic Limit of the M2-brane}

We can use the limit discussed in the previous section to find a consistent non-relativistic M2-brane metric. Let us split our spacetime coordinates into $(t,z,\bar z,u,\bar u,  \vec v)$,  where the transverse coordinate $u$ loosely corresponds to $\calz^1$ and $ \vec v$ to the real and imaginary parts of $\calz^A$ in the field theory. The metric for a stack of M2-branes extended along $(t,z,\bar z)$ at $u = \vec v = 0$ is
\begin{equation} \label{eq: rel metric}
    ds^2 = \hat{\ch}^{-\frac{2}{3}} \brac{-dt^2 + dz d\bar z} + \hat{\ch}^{\frac{1}{3}} \brac{du  d\bar u  + d \vec v\cdot  d \vec v} \ , 
\end{equation}
where the function $\hat{\ch}$ is
\begin{equation}
    \hat{\ch} = 1 + \frac{\hat{R}^6}{\brac{u\bar u +  \vec v\cdot  \vec v}^3} \ .
\end{equation}

Recall that we took the scalings \eqref{eq: scaling 1} and \eqref{eq: scaling 2} for the coordinates and fields in the field theory. We can use a homogeneous conformal transformation\footnote{This is a symmetry of ABJM and will therefore give an equivalent scaling limit.} to put this into the form
\begin{align}
    \brac{t,z, \calz^1 , \calz^A} \to \brac{c t, c^{-\frac{1}{2}} z , c \calz^1, c^{-\inv{2}} \calz^A} \ ,
\end{align}
where the scaling limit takes $c\to\infty$. As we are considering a limit of supergravity in which the Planck length is unchanged, it is natural to use the field theory scaling to postulate the coordinate transformation
\begin{equation} \label{eq: scaling grav}
    \brac{t,z,  u  ,  \vec v} \to \brac{c t, c^{-\frac{1}{2}} z  , c u,  c^{-\inv{2}}  \vec v}
\end{equation}
for the metric \eqref{eq: rel metric}. If we require that $\hat{\ch}$ is non-trivial in the limit $c\to\infty$ then we must also scale the parameter $\hat{R}$ to
\begin{equation}
    \hat{R} = c R \ ,
\end{equation}
where $R$ is taken to be finite and independent of $c$. 
%%%%%%%%%%%%%%%%%%%%%%%%%%%%%%%%%%%%%%%%%%%%%%%%%%%%%%%%%%%
In contrast to the issues discussed in \cite{Avila:2023aey} this limit scales the parameter with a positive power of $c$. Since $R$ is related to the number of M2-branes our solution describes, it appears that our limit requires us to take the number of M2-branes in our theory to infinity. This would suggest that we must also take $N\to\infty$ in the field theory's non-relativistic limit to maintain the duality between the two.  We hope to elaborate on this matter in due course.
%%%%%%%%%%%%%%%%%%%%%%%%%%%%%%%%%%%%%%%%%%%%%%%%%%%%%%%%%%%

After implementing the scaling we find that the metric becomes
\begin{equation}
    ds^2 \to c^2 \bigg[ - \ch^{-\frac{2}{3}} dt^2 + \ch^{\frac{1}{3}} du d\bar u  \bigg] + \inv{c} \bigg[ \ch^{-\frac{2}{3}} dzd\bar z + \ch^{\inv{3}} dv^s d v^s \bigg] \ ,
\end{equation}
where
\begin{equation}
    \ch  = 1 + \frac{R^6}{\brac{u \bar u + c^{-3}  \vec v\cdot  \vec v}^3} \ .
\end{equation}
We see that as $c\to\infty$ the metric becomes a membrane Newton-Cartan structure with
\begin{subequations}
\begin{align}
    \tau_{\mn} dx^{\mu}\otimes dx^{\nu} &= - \brac{1 + \frac{R^6}{\brac{u \bar{u}}^3}}^{-\frac{2}{3}} dt\otimes dt  + \inv{2}\brac{1 + \frac{R^6}{\brac{u \bar{u}}^3}}^{\frac{1}{3}} \brac{du\otimes  d\bar{u} + d\bar{u} \otimes du } \ , \\ 
    H^{\mn} \frac{\partial}{\partial x^{\mu}} \otimes \frac{\partial}{\partial x^{\nu}}  &= 2\brac{1 + \frac{R^6}{\brac{u \bar{u}}^3}}^{\frac{2}{3}} \brac{ \partial \otimes \bar{\partial} + \bar{\partial} \otimes \partial} + \brac{1 + \frac{R^6}{\brac{u \bar{u}}^3}}^{-\frac{1}{3}} \brac{\frac{\partial}{\partial \vec v} \otimes \frac{\partial}{\partial  \vec v} }   \ .
\end{align}
\end{subequations}
From here onwards we will refer to $(t,u,\bar u)$ as $\tau$ coordinates and the rest as $H$ coordinates.

In the near-horizon limit, where we take $R$ to be much larger than any other scale at which we probe the geometry, the spatial and temporal metrics simplify to 
\begin{subequations} \label{eq: non-relativistic ads solution}
\begin{align}
    \tau_{\mn} dx^{\mu}\otimes  dx^{\nu} &= - \frac{\brac{u\bar{u}}^2 dt\otimes dt}{R^4}  + \frac{R^2 }{2u\bar{u}} \brac{du \otimes d\bar{u} + d\bar{u} \otimes du}  \ , \\
    H^{\mn} \frac{\partial}{\partial x^{\mu}} \otimes \frac{\partial}{\partial x^{\nu}} &= \frac{2 R^4}{\brac{u \bar{u}}^2} \brac{ \partial \otimes \bar{\partial} + \bar{\partial} \otimes \partial} + \frac{u \bar{u}}{R^2} \brac{\frac{\partial}{\partial  \vec v} \otimes \frac{\partial}{\partial  \vec v} } \ .
\end{align}
\end{subequations}
The geometry defined by the $\tau$ directions is $AdS_2 \times S^1$, whereas the $H$ geometry is somewhat more exotic and consists of two planes that grow and shrink as $u$ varies. As discussed in the previous section, we can introduce the projective inverses
\begin{subequations}
\begin{align}
    \tau^{\mn} \frac{\partial}{\partial x^{\mu}} \otimes \frac{\partial}{\partial x^{\nu}} &= - \frac{R^4}{\brac{u \bar{u}}^2} \frac{\partial}{\partial t} \otimes \frac{\partial}{\partial t} + \frac{2 u \bar{u}}{R^2} \brac{ \frac{\partial}{\partial u} \otimes \frac{\partial}{\partial \bar{u}} + \frac{\partial}{\partial \bar{u}} \otimes \frac{\partial}{\partial u}} \ , \\
    H_{\mn} dx^{\mu} \otimes dx^{\nu} &= \frac{\brac{u \bar{u}}^2 }{2R^4} \brac{dz \otimes d\bar{z} + d\bar{z} \otimes dz} + \frac{R^2 d \vec v\otimes d \vec v}{u \bar{u}} \ ,
\end{align}
\end{subequations}
though we again note that these are only defined up to the local Galilean boosts \eqref{eq: local galilean boosts}. When we require the use of vielbeins we will take
\begin{subequations}
\begin{align}
    \tau^t &= \frac{u \bar{u}}{R^2} dt \ , \\
    \tau^u &= \frac{R }{u} du \ ,
\end{align}
\end{subequations}
and
\begin{subequations}
\begin{align}
    e^z &= \frac{u^2}{R^2} dz \ , \\
    e^{\vec v} &= \frac{R}{\sqrt{u \bar{u}}} d   \vec v\ .
\end{align}
\end{subequations} 
With this parameterisation the subleading metric fields are
\begin{subequations}
\begin{align}
    m^t &= \frac{\vec{v} \cdot \vec{v}}{R^2} dt \ , \\
    m^u &= - \frac{R \brac{\vec{v} \cdot \vec{v}} }{2 u^2 \bar{u}} du \ , \\
    \pi^z &= \frac{u \brac{\vec{v}\cdot \vec{v}} }{\bar{u} R^2} dz \ , \\
    \pi^{\vec{v}} &= - \frac{R \brac{\vec{v} \cdot \vec{v}} }{2 \brac{u\bar{u}}^{3/2}} d\vec{v} \ .
\end{align}
\end{subequations}

In order for the limiting metric to be a solution of the eleven-dimensional non-relativistic theory we must check that its $C$-field admits a decomposition of the form \eqref{eq: C3 decomp} and that the constraint \eqref{eq: constraint} is satisfied. The $C$-field for the relativistic M2 solution is
\begin{equation}
    \hat{C}_3 = \hat{\ch}^{-1} dt \wedge dx^1 \wedge dx^2  + k
\end{equation}
for some constant 3-form $k$ that will drop out of all physical quantities. Upon taking the scaling \eqref{eq: scaling grav} and the near-horizon limit this becomes
\begin{equation}
    \hat{C}_3 = \frac{i}{2}dt \wedge dz \wedge d\bar z \brac{ \frac{\brac{u \bar{u}}^3}{R^6} + \frac{3 \brac{u \bar{u}}^2  \vec v\cdot  \vec v}{c^3 R^6} + O\brac{\inv{c^6}} } + k \ ,
\end{equation}
which appears to not match \eqref{eq: C3 decomp} due to the absence of a $c^3$ term. However, the volume form in the $\tau$ directions is
\begin{equation}
    \inv{6} \epsilon_{abc} \tau^a\wedge \tau^b \wedge \tau^c= \frac{i}{2}dt \wedge du \wedge d\bar u \ ,
\end{equation}
so taking the constant 3-form to be
\begin{equation}
    k = - \frac{i}{2}c^3 dt \wedge du \wedge d\bar u
\end{equation}
we see that the expansion of $\hat{C}_3$ takes the desired form. It should be noted that, up to an overall sign, $k$ is identical to the background 3-form field \eqref{eq: WZterm} used to cancel the divergent boundary terms in the field theory.  

From this we find that $F_4$ is
\begin{equation} \label{eq: C field strength}
    F_4 =  \frac{3i \brac{u\bar{u}}^2}{2R^6} \brac{u d\bar{u} + \bar{u} du} \wedge dt\wedge dz\wedge d\bar z  \ ,
\end{equation}
so both sides of \eqref{eq: constraint} vanish identically. We therefore expect that the scaled solution solves the equations of motion of the non-Lorentzian theory.  We can also read off the subleading field strength
\begin{equation} \label{eq: C tilde field strength}
    \Tilde{F}_4 = \frac{3i\brac{u \bar{u}}^2}{R^6} \brac{ \vec v\cdot  \vec v \brac{\frac{du}{u} + \frac{d\bar{u}}{\bar{u}}} +  \vec v\cdot  d  \vec v} \wedge dt \wedge dz\wedge d\bar z \ .
\end{equation}
We can compute the on-shell value of the Lagrange multiplier field $G_{I_1 ... I_4}$ from the expressions for the subleading fields using \eqref{eq: on-shell G}, where we find
\begin{equation}
    G_{I_1 ... I_4} = 0 \ .
\end{equation}

Unfortunately, the supersymmetric extension of the Bosonic theory presented in \cite{Blair:2021waq} is not known, so we can't discuss whether our solution preserves any supercharges from the non-Lorentzian perspective. In a way, the conjecture that there is still a holographic duality between the near-BPS limit field theory discussed in section \ref{sect: near bps limit} and a prospective eleven-dimensional membrane Newton-Cartan supergravity theory in asymptotically AdS spacetimes (i.e. the asymptotic behaviour of the metric resembles \eqref{eq: non-relativistic ads solution}) is a prediction that the supersymmetric completion of the gravity theory exists, and that the metric structures \eqref{eq: non-relativistic ads solution} and 4-form field strength \eqref{eq: C field strength} form a maximally supersymmetric solution of the theory.

\subsection{Gravitational Symmetries}

\subsubsection{Isometries of the Near-Horizon Geometry}

An infinitesimal coordinate transformation
$\delta x^{\mu} = \xi^{\mu}$ is an isometry of the Newton-Cartan structure if both $\tau$ and $H$ are unchanged up to local invariances. Recalling that these are the local Galilean boosts and dilatations, the transformations need to obey the equations
\begin{subequations} \label{eq: NC killing}
\begin{align} \label{eq: NC killing a}
    0 &=  \cal{L}_{\xi} \tau_{\mn} + 2 \Lambda \tau_{\mn} \ , \\ \label{eq: change in tau}
    0 &= \cal{L}_{\xi} \tau^{\mn} - 2 \tau_a^{(\mu} H^{\nu) \rho} \lambda_I^a e^I_{\rho} - 2 \Lambda \tau^{\mn} \ ,
\end{align}
involving $\tau$, and the equations
\begin{align} \label{eq: NC Killing b}
    0 &= \cal{L}_{\xi} H^{\mn} + \Lambda H^{\mn} \ , \\  \label{eq: change in H}
    0 &= \cal{L}_{\xi} H_{\mn} + 2 \eta_{ab} \lambda^a_I e^I_{(\mu} \tau^b_{\nu)} - \Lambda H_{\mn} \ ,
\end{align}
\end{subequations}
involving $H$. 

Let's focus for now on \eqref{eq: NC killing a}. The mixed $\tau$ and $H$ coordinate components just impose that $\xi^{t}$ and $\xi^{u}$ are independent of the $H$ coordinates, and the $uu/\bar{u}\bar{u}$ components impose that $\xi^u$ cannot depend on $\bar{u}$. The system of equations that we have to solve is then
\begin{subequations}
\begin{align}
    \inv{u} \xi^u + \inv{\bar{u}} \xi^{\bar{u}} &= \inv{3} \brac{\partial_u \xi^u + \partial_{\bar{u}} \xi^{\bar{u}} - 2 \partial_t \xi^t} \ , \\
    \partial_t \xi^u &= \frac{2\brac{u \bar{u}}^3}{R^6} \partial_{\bar{u}} \xi^t \ , \\
    \Lambda &= -\inv{3} \brac{\partial_t \xi^t + \partial_u \xi^u + \partial_{\bar{u}} \xi^{\bar{u}}} \ .
\end{align}
\end{subequations}
A bit of work shows that the solution to this is
\begin{subequations}
\begin{align}
    \xi^t &= a + b t + c t^2 + \frac{c R^6}{4\brac{u \bar{u}}^2} - \frac{R^6}{4\bar{u}^2} \brac{\beta + 2 \gamma t } - \frac{R^6}{4 u^2} \brac{\bar{\beta} + 2 \bar{\gamma} t}   \ , \\
    \xi^{u} &= - \frac{b}{2} u - c u t + u^3 \brac{\alpha + \beta t + \gamma t^2} + \frac{R^6 \bar{\gamma}}{4 u} + i \theta u \ , \\
    \Lambda &= -\brac{\alpha + \beta t + \gamma t^2} u^2 - \brac{\bar{\alpha} + \bar{\beta} t + \bar{\gamma} t^2} \bar{u}^2 + \frac{R^6 \bar{\gamma}}{4 u^2} + \frac{ R^6 \gamma}{4 \bar{u}^2} \ ,
\end{align}
\end{subequations}
in terms of real infinitesimal parameters $\{a,b,c,\theta\}$ and complex infinitesimal parameters $\{\alpha,\beta,\gamma\}$. The algebra the transformations form is isomorphic to $\frak{so}(2,3)$, which can be seen by noting that   $\tau$ is conformally equivalent to the flat metric on  $\bb{R}^{1,2}$ and the solutions to \eqref{eq: NC killing a} are the conformal Killing vectors of $\tau$.

We can work through \eqref{eq: NC Killing b} or \eqref{eq: change in H} in a similar way. The only difference between the two arises in the purely $\tau$ and mixed component equations. In \eqref{eq: NC Killing b}, the purely $\tau$ components vanish and the mixed components impose the same constraint found in \eqref{eq: NC killing a} that $\{\xi^t, \xi^u, \xi^{\bar{u}}\}$ must be independent of the $H$ coordinates. In \eqref{eq: change in H}, the purely $\tau$ components of the equation impose that the $\tau$ components of the 1-forms $\lambda^a$ vanish. The mixed components fix the form of the Galilean boosts required for the transformations found from \eqref{eq: NC killing a} and allow for any new constants of integration introduced in $\{\xi^z,\vec \xi\}$ to be given arbitrary dependence on $(t,u,\bar{u})$ through an appropriate choice of the aforementioned 1-forms. This is consistent with \eqref{eq: NC Killing b} as in that formulation any constants introduced in the solution of these equations alone are never subject to derivatives along the $\tau$ directions, so their dependence on these is arbitrary.

The remaining equations to be solved are
\begin{subequations}
\begin{align}
    \partial_z \xi^z + \partial_{\bar{z}} \xi^{\bar{z}} &= -2 \brac{\inv{u} \xi^u + \inv{\bar{u}} \xi^{\bar{u}}} + \Lambda \ , \\
    \partial_{\bar{z}} \xi^z &= 0 \ , \\
    \partial_r\xi^s + \partial_s \xi^r &= \delta_{rs} \brac{\inv{u} \xi^u + \inv{\bar{u}} \xi^{\bar{u}} + \Lambda} \ , \\
    2 \partial_{\bar{z}} \xi^r + \brac{\frac{u \bar{u}}{R^2}}^3 \partial_r \xi^z &= 0 \ .
\end{align}
\end{subequations}
where we have introduced indices  $r,s=1,2,3,...,6$ for vectors in ${\mathbb R}^6$; $\vec x \to \xi^r$, $\vec \xi \to \xi^r$.
The solutions to these can be taken to be\footnote{There is some ambiguity in the split of some terms between $\xi^z$ and $\xi^{\bar{z}}$, but this just corresponds to a particular parameterisation of the function $r$.} 
\begin{subequations} \label{eq: spatial killing comp}
\begin{align}
    \xi^z &= \brac{b + 2ct  - 3 u^2 \brac{\alpha + \beta t + \gamma t^2} - \frac{R^6 \bar{\gamma}}{4 u^2}} z  + \chi + 2 v^a \rho^a + i r z \ , \\
    \xi^r &= \brac{\frac{R^6 \bar{\gamma}}{4 u^2} + \frac{R^6 \gamma}{4 \bar{u}^2} - \frac{b}{2} - ct} v^r + \tensor{R}{^r_s} v^s + k^r + \brac{\frac{u \bar{u}}{R^2}}^3 \brac{\bar{z} \rho^r + z \bar{\rho}^r} \ ,
\end{align}
\end{subequations}
where $\{\chi,\vec \rho\}$ are complex, $\{r,R,k\}$ are real, $R$ is antisymmetric, and all are arbitrary functions of $(t,u,\bar{u})$. However, to make contact with the field theory we must implement a $\bb{Z}_k$ orbifold of the $\bb{R}^6$ parameterised by $\{\vec v\}$ as discussed in appendix \ref{sect: orbifold}. Requiring that the Killing vectors are globally defined on the orbifold forces us to take
\begin{equation}
    \vec k = \vec\rho = 0 \ ,
\end{equation}
which we shall assume from here onwards, and restricts the allowed rotation matrices $R$ to an $\frak{su}(3)\oplus \frak{u}(1)$ subalgebra. We can then read off the required local Galilean boosts from \eqref{eq: change in H}, where we find
\begin{subequations} \label{eq: boosts for transf}
\begin{align}
    \lambda^t_z &= \frac{\bar{u}}{2 u} \Big[ \bar{z} \brac{2c - 3 \bar{u}^2 \brac{\bar{\beta} + 2 \bar{\gamma} t} - i \partial_t r} + \partial_t \bar{\chi} \Big] \ , \\
    \lambda^t_r &= \frac{R^3}{\brac{u \bar{u}}^{\frac{3}{2}} } \Big[ \partial_t \tensor{R}{^r_s} v^s - c v^r \Big] \ , \\
    \lambda^u_z &= \frac{\bar{u}^3}{R^3} \bigg[ \bar{z} \brac{6 \bar{u} \brac{\bar{\alpha} + \bar{\beta} t + \bar{\gamma} t^2} + i \partial_{\bar{u}}r - \frac{R^6 \gamma}{2 \bar{u}^3}} - \partial_{\bar{u}} \bar{\chi} \bigg] \ , \\
    \lambda^{\bar{u}}_z &=  \frac{u \bar{u}^2}{R^3} \Big[ i \bar{z} \partial_u r - \partial_u \bar{\chi} \Big] \ , \\
    \lambda^u_r &= \sqrt{\frac{\bar{u}}{u}} \bigg[ \frac{R^6 \gamma}{\bar{u}^3} v^r - 2 \partial_{\bar{u}} \tensor{R}{^r_s} v^s \bigg] \ .
\end{align}
\end{subequations}

Finally, we must check that everything computed above is consistent with \eqref{eq: change in tau}. The components of the equation vanish when both $\mu$ and $\nu$ are spatial indices, and when both are temporal the equation is automatically satisfied as $\tau^{\mn}$ is the inverse of $\tau_{\mn}$ when projected onto the temporal coordinates. The only non-trivial check that must be performed is therefore for the mixed components, for which the equation is
\begin{equation}
    \partial_{\nu} \xi^r \tau^{\mu\nu} + \tau_a{}^{\mu} H^{rs} e_{s}^I \lambda^a_I = 0 \ .
\end{equation} 
It is straightforward to check that every component of this is satisfied for the spatial Killing vector components \eqref{eq: spatial killing comp} and local boosts \eqref{eq: boosts for transf}.

\subsubsection{Form-Field Symmetries}
So far we have only determined which coordinate transformations are isometries of the metric structures. However, this is not enough to determine whether we have a symmetry of the solution; the 3-form field $C_3$ and $SO(8)$ 4-form field $G_4$ must also be invariant. The transformation of the fields under the isometries given by the vector field $\xi^{\mu}$ and local boosts $\lambda^a_I$ are given by
\begin{subequations}
\begin{align}
    \delta C_3 &= \cal{L}_{\xi}C_3 - \inv{2} \epsilon_{abc} \lambda^a_I e^I \wedge \tau^b \wedge \tau^c + d \sigma_2 \ , \\ \label{eq: tilde F transformation}
    \delta G_{I_1 ... I_4} &= \cal{L}_{\xi} G_{I_1 ... I_4}- 4 \lambda^a_{[I} e^{\mu}_{J} e^{\nu}_K e^{\rho}_{L]} \tau^{\sigma}_a F_{\mu \nu\rho\sigma} \ .
\end{align}
\end{subequations}
Note that by including $\sigma_2\in\Omega^{2}(\cal{M})$ we allow for arbitrary gauge transformations, which are of course a local symmetry of the theory. The symmetries of the solution are then determined by the conditions $\delta C_3 = \delta G_{I_1 ... I_4} = 0$.

We will deal with $C_3$ first. The transformations parameterised by $\{a,b,\theta \}$ all leave $C_3$ invariant and aren't associated with any local boosts, so these isometries are full symmetries of our solution. The transformation parameterised by $c$ is non-trivial, but the new terms in $\delta C_3$ are exact (and can therefore be absorbed by a particular choice of $\sigma_2$) so this is also a symmetry. Things get more interesting when we consider the transformation parameterised by $\alpha$. The Lie derivative and local transformations are given by
\begin{subequations}
\begin{align}
    \cal{L}_{\xi} C_3 &= - \frac{3i \brac{u \bar{u}}^3 }{R^6} \brac{\alpha u z dt \wedge du \wedge d\bar{z} + \bar{\alpha} \bar{u} \bar{z} dt \wedge dz \wedge d\bar{u}} \ , \\
    \epsilon_{abc} \lambda^a_I e^I \wedge \tau^b \wedge \tau^c &= \frac{6i (u \bar{u})^3}{R^6} \brac{\alpha u z dt \wedge du \wedge d\bar{z} + \bar{\alpha} \bar{u} \bar{z} dt \wedge dz \wedge d\bar{u} } \ ;
\end{align}
\end{subequations}
as these do not cancel and are not exact we see that $\delta C_3 \neq 0$, so $\alpha$ is not a symmetry of $C_3$. Similarly, the $\beta$ transformation induces the terms
\begin{subequations}
\begin{align} \nonumber
    \cal{L}_{\xi} C_3 = \, &- \frac{3i t \brac{u \bar{u}}^3}{R^6} \brac{\beta z u dt \wedge du \wedge d\bar{z} + \bar{\beta} \bar{z} \bar{u} dt \wedge dz \wedge d\bar{u} } \\
    & + \frac{R^6}{2} \brac{\frac{\bar{\beta} du}{u^2} + \frac{\beta d \bar{u}}{\bar{u}^3}} \wedge dz \wedge d\bar{z} \ , \\ \nonumber
    \epsilon_{abc} \lambda^a_I e^I \wedge \tau^b \wedge \tau^c = \, &\frac{6i t \brac{u \bar{u}}^3}{R^6} \brac{\beta z u dt \wedge du \wedge d\bar{z} + \bar{\beta} \bar{z} \bar{u} dt \wedge dz \wedge d\bar{u} }  \\
    & - \frac{3i}{2} \brac{\beta z u^2 d \bar{z} + \bar{\beta} \bar{z} \bar{u}^2 dz} \wedge du \wedge d\bar{u} \ ,
\end{align}
\end{subequations}
while $\gamma$ induces
\begin{subequations}
\begin{align} \nonumber
    \cal{L}_{\xi} C_3 = \, & \frac{it}{2} \brac{\gamma u^3 d\bar{u} + \bar{\gamma} \bar{u}^3 du} \wedge dz \wedge d\bar{z} \\
    &+ \frac{i dt}{4}\wedge \brac{\gamma u^3 \bar{z} dz \wedge d\bar{u} + \bar{\gamma} \bar{u}^3 z du \wedge d\bar{z} } \ , \\ \nonumber
    \epsilon_{abc} \lambda^a_I e^I \wedge \tau^b \wedge \tau^c = \, & - 3it \brac{\gamma u z d\bar{z} + \bar{\gamma} \bar{u} \bar{z} dz} \wedge du \wedge d\bar{u} \\ \nonumber
    &+ \frac{i \brac{u \bar{u}}^3 \bar{z}}{R^6} \brac{6 t \bar{u} \bar{\gamma} - \frac{R^6 \gamma}{2 \bar{u}^3}} dt \wedge dz \wedge d\bar{u} \\
    &+ \frac{i \brac{u \bar{u}}^3 z}{R^6} \brac{6t u \gamma - \frac{R^6 \bar{\gamma}}{2u^3}} dt \wedge du \wedge d\bar{z} \ ,
\end{align}
\end{subequations}
so neither of these are symmetries of $C_3$.

What about the transformations acting solely on the $H$ coordinates? Recalling that the metric transformations \eqref{eq: NC killing} allowed $\{\chi, r, \tensor{R}{^r_s}\}$ to have arbitrary dependence on the $\tau$ coordinates, the transformation of $C_3$ is
\begin{subequations}
\begin{align} \nonumber
    \cal{L}_{\xi} C_3 = \, & \frac{i \brac{u \bar{u}}^3}{2 R^6} \bigg( \partial_u \chi dt \wedge du \wedge d\bar{z} + \partial_{\bar{u}} \chi dt \wedge d \bar{u} \wedge d\bar{z} \\
    &+ \partial_u \bar{\chi} dt \wedge dz \wedge du + \partial_{\bar{u}} \bar{\chi} dt \wedge dz \wedge d \bar{u} \bigg) \ , \\ \nonumber
    \epsilon_{abc} \lambda^a_I e^I \wedge \tau^b \wedge \tau^c = \,& \frac{i}{2} \partial_t \brac{ \bar{\chi} dz +  \chi d\bar{z}} \wedge du \wedge d\bar{u} \\
    &- \frac{i \brac{u \bar{u}}^3}{R^6} \bigg( \partial_{\bar{u}} \brac{\bar{\chi} dz +  \chi d \bar{z}} \wedge d\bar{u} \wedge dt \\ \nonumber
    &+ \partial_u \brac{\bar{\chi} dz +  \chi d \bar{z}} \wedge dt \wedge du \bigg ) \ ,
\end{align}
\end{subequations}
for $\chi$,
\begin{subequations}
\begin{align} \nonumber
    \cal{L}_{\xi} C_3 = \, & - \frac{ \brac{u \bar{u}}^3}{2 R^6} \bigg(  \partial_u r dt \wedge du \wedge d\brac{z\bar{z}} -  \partial_{\bar{u}} r dt  \wedge d\brac{z\bar{z}}\wedge d \bar{u} \bigg) \ , \\ \nonumber
    \epsilon_{abc} \lambda^a_I e^I \wedge \tau^b \wedge \tau^c = \,& \frac{1}{2} \partial_t r \brac{ \bar{z} dz - z d\bar{z}} \wedge du \wedge d\bar{u} \\
    &- \frac{ \brac{u \bar{u}}^3}{R^6} \bigg( \partial_{\bar{u}}r \brac{\bar{z} dz -z d \bar{z}} \wedge d\bar{u} \wedge dt \\ \nonumber
    &+ \partial_u r \brac{\bar{z} dz - z d \bar{z}} \wedge dt \wedge du \bigg ) \ ,
\end{align}
\end{subequations}
for $r$, and 
\begin{subequations}
\begin{align}
    \cal{L}_{\xi} C_3 =\, & 0 \ , \\ \nonumber
    \epsilon_{abc} \lambda_I^a e^I \wedge \tau^b \wedge \tau^c = \, & \frac{i R^6}{\brac{u \bar{u}}^3} \partial_t \tensor{R}{^r_s} v^s dv^r\wedge du \wedge d\bar{u} \\
    & - 2i v^s dv^r \wedge dt \wedge \brac{ \partial_u \tensor{R}{^r_s} du - \partial_{\bar{u}} \tensor{R}{^r_s} d\bar{u}} \ ,
\end{align}
\end{subequations}
for $\tensor{R}{^r_s}$. We see that these are only symmetries of $C_3$ if the functions are taken to be constants. The symmetries of $C_3$ are then the transformations with constant parameters $\{a,b,c,\theta, \chi, r, \tensor{R}{^r_s} \}$. It is straightforward to check that \eqref{eq: tilde F transformation} vanishes for each of these, so the symmetry algebra of the gravitational solution is
\begin{equation} \label{eq: grav symmetries}
    \frak{g} = \frak{so}(1,2) \oplus \frak{u}(1)_{\theta} \oplus \frak{iso}(2) \oplus \frak{su}(3) \oplus \frak{u}(1)_R \ .
\end{equation}

\subsubsection{Boundary Structure and Asymptotic Symmetries} \label{sect: asymptotic sym}

In the previous section we found the exact symmetries of the non-relativistic eleven-dimensional background. However, in the holographic context we expect to match global symmetries of our field theory with the asymptotic symmetries of the gravity dual, that is diffeomorphisms which only need to be symmetries as we approach the boundary. While this necessarily contains the symmetries previously discussed, there is the potential for an extended symmetry group to emerge once these are taken into account. These can either take the form of new transformations that leave the Newton-Cartan metric structures and 3-form field invariant at leading order near the boundary, or exact isometries of the metric structure that are only symmetries of the 3-form field asymptotically.

To discuss the boundary in the (relativistic) $AdS_4$ solution we can go to the Poincar\'e patch with radial coordinate $\hat \rho$  defined by the coordinate transformation
\begin{equation}
    \hat{\rho} = \frac{\hat{R}^3}{2\hat{r}^2} \ ,
\end{equation}
so that the boundary lies at $\hat\rho=0$. In our coordinates  we have 
\begin{equation}
    \hat{\rho} = \frac{c R^3}{2\brac{u^2 + \frac{\vec v\cdot\vec v}{c^3}}} \ .
\end{equation}
We can define analogous coordinates for the $AdS_2$ in $\tau$ and radial coordinate in $H$ by
\begin{subequations}
\begin{align}
   \rho &= \frac{R^3}{2  u\bar{u}} \ , \\
   \sigma &= \frac{R^3}{2  \vec v\cdot\vec v } \ ,
\end{align}
\end{subequations}
so
\begin{equation}
    \hat{\rho} = \frac{c \rho \sigma}{\sigma + \inv{c^3} \rho}   \ .
\end{equation}
At finite $c$ the $AdS_4$ boundary $\hat{\rho}=0$ corresponds to $\rho=0$ or $\sigma=0$. However in the non-relativistic limit  $c\to\infty$ we have, for any non-zero $\sigma$, $\hat\rho = c\rho$. Thus  to match the isometries found in the previous section onto the field theory transformations we   take the boundary to be $\rho = 0$, which is a subset of the original boundary.   The metric structures and 3-form field in this coordinate system are
\begin{subequations} \label{eq: boundary coordinate system}
\begin{align}
    \tau_{\mn} dx^{\mu}\otimes   dx^{\nu} &= \frac{R^2 \brac{-dt\otimes dt  + d\rho\otimes d\rho}}{4\rho^2} + R^2 d\theta\otimes d\theta \ , \\
    H^{\mn} \frac{\partial}{\partial x^{\mu}} \otimes \frac{\partial}{\partial x^{\nu}} &= \frac{8 \rho^2}{R^2} \brac{\partial \otimes \bar{\partial} + \bar{\partial} \otimes \partial} + \frac{4\sigma^3}{\rho R^2}    \frac{\partial}{\partial \sigma} \otimes \frac{\partial}{\partial \sigma} + \frac{\sigma}{\rho R^2}g^{-1}_{S^5 / \bb{Z}_k}  \ , \\
    C_3 &= \frac{i R^3}{16 \rho^3} dt \wedge dz \wedge d\bar{z} \ ,
\end{align}
\end{subequations}
so if we require that the radius of the $S^5/\bb{Z}_k$ factor remains finite at the boundary we should also take $\sigma$ to zero with the ratio of $\rho$ and $\sigma$ held fixed. However, these considerations won't be relevant to our discussion. We note in passing that as we approach the boundary the co-metric in the $z\bar{z}$ plane vanishes, which could be a bulk indication of the 'gauging' of spatial transformations in the field theory; we will not pursue this point any further in this work.

We do not possess a complete understanding of the appropriate boundary conditions ({\it i.e.} the fall-off required in the subleading terms) for a solution of the theory in \cite{Blair:2021waq} to be asymptotically of the form we've presented, so we cannot say with certainty when a transformation is an asymptotic symmetry of our solution. We will therefore be somewhat heuristic and demand that each component of the change in our fields is subleading in $\rho$ with respect to the components \eqref{eq: boundary coordinate system} in the boundary coordinate system, since this seems likely to be a necessary condition for the system to asymptotically approach the required form.

As mentioned previously, there are two classes of potential asymptotic symmetries we could consider. The first are transformations that are exact symmetries of the metric structures but not of the 3-form field but which may enhance to symmetries at the boundary. However, upon examining the transformations of $C_3$ for each isometry that don't form symmetries we see that every transformation contains terms that aren't subleading with respect to the original components as we approach the boundary. We can therefore rule out this class of asymptotic symmetries.

The second are new transformations that asymptotically preserve the metric, which we may hope also are symmetries of $C_3$ at the boundary. While we won't be exhaustive in our discussion of these, there is a particularly important case we must address. In \cite{Bagchi:2009my} it was proposed that the duals of theories invariant under an infinite-dimensional extension of the Galilean conformal algebra should be Newton-Cartan geometries with an $AdS_2$ factor in $\tau$\footnote{The authors of that work were interested solely in non-relativistic limits of string theory, so $\tau$ was taken to only have two non-zero eigenvalues. However, since our $\tau$ geometry factorises into $AdS_2\times S^1$ the discussion will be similar.}, with the infinite-dimensional extension arising from the asymptotic symmetries of the $AdS_2$ metric \cite{Hotta:1998iq, Cadoni:1999ja}. Such infinite dimensional symmetries are not present in our field theory, we would like to see what goes wrong with this argument in our system. Let us examine an adapted version of their argument for these symmetries for our geometry. We observe that the infinitesimal transformation
\begin{subequations}
\begin{align}
    t' &= t + f(t) + \frac{R^6 f''}{8 \brac{u \bar{u}}^2} \ , \\
    u' &= u \brac{1 - \frac{f'}{2}} \ ,
\end{align}
\end{subequations}
changes $\tau$ to
\begin{equation}
    \tau' = - \frac{\brac{u \bar{u}}^2}{R^4} \brac{1 + \frac{R^6 f'''}{4 \brac{u \bar{u}}^2}} dt \otimes dt + \frac{R^2 }{2 u \bar{u}} \brac{du \otimes d\bar{u} + d\bar{u} \otimes du} \ ,
\end{equation}
leaving the geometry invariant up to terms that are subleading as we approach the boundary. Combining the transformation of $u$ with
\begin{subequations}
\begin{align}
    z' &= z \brac{1 + f'} \ , \\
    \vec v^{ \prime} &= \vec v \brac{1 - \frac{f'}{2}} \ ,
\end{align}
\end{subequations}
leaves $H$ invariant, provided we take the local Galilean boost to be
\begin{subequations}
\begin{align}
    \lambda^t_z &= \frac{\bar{u} \bar{z} f''}{2 u} \ , \\
    \lambda^t_r &= - \frac{R^3 f'' v^r}{2 \brac{u \bar{u}}^{\frac{3}{2}}} \ .
\end{align}
\end{subequations}
With these, a short computation gives the contributions
\begin{subequations}
\begin{align}
    \cal{L}_{\xi} C_3 &= \frac{i \brac{u \bar{u}} f'''}{2} dt \wedge dz \wedge d\bar{z} - \frac{i f''}{8} d\brac{u \bar{u}} \wedge dz \wedge d \bar{z} \ , \\
    \epsilon_{abc} \lambda^a_I e^I \wedge \tau^b \wedge \tau^c &= \frac{i f''}{2} \brac{dz + d \bar{z}} \wedge du \wedge d\bar{u} - \frac{i R^6 f''  }{2 \brac{u \bar{u}}^3}\vec v\cdot  d\vec v \wedge du \wedge d\bar{u} \ ,
\end{align}
\end{subequations}
to the transformation of $C_3$. After rewriting this in terms of $\rho$ we see the transformation induces new terms in $C_3$ that diverge faster than the background solution    as we take $\rho\to0$. Thus it is reasonable to assert that this is not an asymptotic symmetry of the theory.

As we find no additional symmetries, we propose that the asymptotic symmetry algebra is \eqref{eq: grav symmetries}. Aside from the $\frak{iso}(2)$ factor, these coincide with the rigid symmetries of the field theory. At first glance the $\frak{iso}(2)$ factor appears not to match with the enhanced spatial conformal symmetry we saw there. However, since we interpret the field theory symmetries as redundancies this is perfectly natural- we're only interested in matching the symmetries associated with charges that act non-trivially on the phase space across both sides of the duality. From this perspective, the oddity is the appearance of the factor of $\frak{iso}(2)$. For our conjectured duality to hold, the charges generated by these Killing vectors
should vanish. Showing this goes beyond the scope of this paper, and is a topic for future work.

\section{Conclusion} \label{sect: conclusion}

In this paper we have discussed a peculiar non-relativistic limit of the Chern-Simons-matter theory associated to M2-branes. The resulting theory has been constructed and analysed before in \cite{Lambert:2018lgt, Lambert:2019nti,Kucharski:2017jwv}, where it was shown to maintain the same number of supersymmetries as the parent theory and its dynamics reduces to motion on a Hitchin moduli space. Here we saw that the spatial symmetry algebra of the theory is an infinite-dimensional enhancement of the algebra one would naively expect to obtain from a non-relativistic limit of ABJM. However, as the associated charge reduces to a boundary term we interpret these transformations as redundancies of the theory. We also considered the corresponding limit taken in the dual eleven-dimensional supergravity theory. This turned out to be a special case of the general membrane-Newton-Cartan limit given in \cite{Blair:2021waq}, giving us a solution in which the $AdS_4$ factor in the relativistic theory is reduced to an $AdS_2$. The symmetries of the solution were calculated and found to contain the physical field theory symmetries, leading us to propose that the duality between the two theories remains after taking the limit on both sides.

There are some outstanding questions that require further study. In particular it is important to better understand the boundary in the dual AdS geometry. We argued above that the field theory resides at the $\rho=0$ boundary of $AdS_2$. While this is a subset of the original $AdS_4$ boundary, we have effectively ignored the section of the boundary at $\sigma = 0$. As this lies in $H$ and not $\tau$ the interpretation of this is not clear. On a related note, unlike in well-understood AdS duals the non-field theoretic part of the $H$ geometry is non-compact and singular as we approach the $\rho\to 0$ boundary. Since we have discussed a limit of a well-defined AdS/CFT pair we expect that the pathologies associated to this can be alleviated. For instance, a prescription where we consider the boundary $\rho\to 0$ and $\sigma \to 0$ with their ratio fixed would answer both this question and the one previously discussed, though it is not obvious that this is the correct thing to consider. It is desirable to have a clearer understanding of these matters. Another issue is the physical role of the $\frak{iso}(2)$ algebra symmetry of the gravitational solution. As discussed in section \ref{sect: asymptotic sym}, for the proposed duality to hold these transformations should not lead to an action on the phase space of the theory; a calculation determining whether this is true, as well a broader discussion of charges in Newton-Cartan-type gravity theories, is therefore of considerable interest.

We close by discussing some future directions it may be fruitful to pursue. In the regime $k^5\gg N$ M-theory on $AdS_4\times S^7/{\mathbb Z}_k$ backgrounds reduces to type IIA string theory on $AdS_4\times {\mathbb CP}^3$ backgrounds, for which there is a known exact worldsheet CFT \cite{Arutyunov:2008if,Stefanski:2008ik}. It is therefore reasonable to ask if there is a non-relativistic limit of the string $\sigma$-model that is related to the large $k$ limit of the MNC solution. Though we have only discussed a single AdS/CFT pairing in this work, we also expect there to be similar non-relativistic limits for other dualities. The obvious example to discuss would be a limit of $\cal{N}=4$ super Yang-Mills that isolates BPS field configurations, along with a corresponding limit of type IIB supergravity on $AdS_5\times S^5$ backgrounds. As this is closely related to the Spin Matrix Theory limits of $\cal{N}=4$ \cite{Harmark:2014mpa} it would be illuminating to see if there is a link between the two approaches.

\section*{Acknowledgements}

We would like to thank Dionysios Anninos and Chris Blair for interesting discussions. N.L. is supported in part  by the STFC consolidated grant ST/X000753/1. J.S. is supported by the STFC studentship ST/W507556/1.

\appendix

\section{Fermions} \label{sect: fermions}

\subsection{Fermion Action and Symmetries} \label{sect: fermion symmetries}

The Bosonic action \eqref{eq: bosonic action} admits a supersymmetric completion \cite{Lambert:2019nti} that arises by taking the non-relativistic limit of the Fermion terms in the ABJM action: let us review how this works. To find the correct limit we must split the spinor fields into chiral components with respect to the matrix
\begin{equation}
    \Gamma = i \gamma^0 \ ,
\end{equation}
where $\{\gamma^{\mu}\}$ are chosen to be real 2x2 matrices satisfying $\gamma^0 \gamma^1 \gamma^2 = \bbm{1}$. In other words, using the projection operator $P_{\pm} = \inv{2} \brac{\bbm{1} \pm \Gamma}$ we define
\begin{equation}
    \hat{\psi}_M^{\pm} = P_{\pm} \hat{\psi}^M \ .
\end{equation}
As our spinors initially had 2 components, the chiral fields are (single-component) complex Grassmann-valued fields. It will be convenient to work with the complex-coordinate gamma matrices
\begin{equation}
    \gamma_z = \inv{2} \brac{\gamma_1 - i \gamma_2} \ ,
\end{equation}
and its conjugate. We see that these satisfy
\begin{subequations}
\begin{align}
    \gamma_z \Gamma &= \gamma_z \ , \\
    \gamma_{\bar{z}} \Gamma &= - \gamma_{\bar{z}} \ ,
\end{align}
\end{subequations}
so $\gamma_z$ annihilates spinors of negative chirality (and similarly for $\gamma_{\bar{z}}$). Putting this all together, we can rewrite the ABJM Fermion terms as
\begin{align} \nonumber
    \hat{S}_F = \tr \int d^3\hat{x} \bigg( & i \hbpsi^{M,+} \hd_0 \hpsi_M^+ + i \hbpsi^{M,-} \hd_0 \hpsi^-_M - 2i \hbpsi^{M,+} \hd \hpsi_M^- - 2i \hbpsi^{M,-} \hbd \hpsi_M^+ \\ \nonumber
    &- \frac{2\pi}{k} \hbpsi^{M,+} [\hpsi_M^+, \hcalz^N ; \hbcalz_N] + \frac{2\pi}{k} \hbpsi^{M,-} [\hpsi_M^-, \hcalz^N ; \hbcalz_N] + \frac{4\pi}{k} \hbpsi^{M,+} [\hpsi_N^+ , \hcalz^N ; \hbcalz_M] \\ \nonumber
    &- \frac{4\pi}{k} \hbpsi^{M,-} [\hpsi_N^- , \hcalz^N ; \hbcalz_M] + \frac{\pi}{k} \epsilon_{MNPQ} \hbpsi^{M,+} [ \hcalz^P , \hcalz^Q ; \hbpsi^{N,-}] \\ \nonumber
    &- \frac{\pi}{k} \epsilon_{MNPQ} \hbpsi^{M,-} [ \hcalz^P , \hcalz^Q ; \hbpsi^{N,+}] - \frac{\pi}{k} \epsilon^{MNPQ} \hpsi^+_M [ \hbcalz_P , \hbcalz_Q ; \hpsi^-_N] \\
    &+ \frac{\pi}{k} \epsilon^{MNPQ} \hpsi^-_M [\hbcalz_P, \hbcalz_Q ; \hpsi_N^+]
    \bigg) \ ,
\end{align}
where $\hbpsi^{M, \pm}$ denotes the Hermitian conjugate of $\hpsi_M^{\pm}$. Under the scaling \eqref{eq: scaling 1}, we take our Fermions to transform as
\begin{subequations}
\begin{align}
    \hpsi^{1,-}(\hat{t},\hat{z}, \hat{\bar{z}}) &= \inv{\omega} \psi^{1,-}(t,z,\bar{z}) \ , \\
    \hpsi^{1,+}(\hat{t},\hat{z}, \hat{\bar{z}}) &= \psi^{1,+}(t,z,\bar{z}) \ , \\
    \hpsi^{A,-}(\hat{t},\hat{z}, \hat{\bar{z}}) &= \psi^{A,-} (t,z,\bar{z}) \ , \\
    \hpsi^{A,+}(\hat{t},\hat{z}, \hat{\bar{z}}) &= \inv{\omega} \psi^{A,+}(t,z,\bar{z}) \ .
\end{align}
\end{subequations}
Combining this with the transformation \eqref{eq: scaling 2} of the Bosonic fields leads to a finite action, so we can take the limit $\omega\to0$ and find the Fermion action
\begin{align} \label{eq: fermionic action} \nonumber
    S_F = \tr \int d^3x \bigg[ & i \bpsi^{1,-} D_0 \psi_1^- - 2i \bpsi^{1,+}  D \psi_1^- - 2i \bpsi^{1,-} \bar{D} \psi_1^+ + i \bpsi^{A,+} D_0 \psi^+_A \\ \nonumber
    &- 2i \bpsi^{A,+}  D \psi_A^- - 2i \bpsi^{A,-} \bar{D} \psi_A^+ + \frac{2\pi }{k} \Big( 2\bpsi^{1,+} [\psi_1^+, \calz^1 ; \bcalz_1] \\ \nonumber
    &+ \bpsi^{1,-} [\psi_1^-, \calz^A ; \bcalz_A] +2 \bpsi^{A,-} [\psi_A^-, \calz^1 ; \bcalz_1] - \bpsi^{A,+} [\psi_A^+, \calz^B ; \bcalz_B] \\ \nonumber
    & + 2 \bpsi^{1,+} [\psi_A^+, \calz^A ; \bcalz_1]  - 2 \bpsi^{1,-} [\psi_A^-, \calz^A ; \bcalz_1] + 2 \bpsi^{A,+} [\psi_1^+, \calz^1 ; \bcalz_A] \\ \nonumber
    &- 2 \bpsi^{A,-} [\psi_1^-, \calz^1 ; \bcalz_A] + 2 \bpsi^{A,+} [\psi_B^+ , \calz^B ; \bcalz_A] \Big) \\
    & + \frac{4\pi}{k} \Big( \epsilon_{ABC} \bpsi^{A,+} [\calz^B, \calz^1 ; \bpsi^{C,-}] - \epsilon^{ABC} \psi^+_A [ \bcalz_B, \bcalz_1 ; \psi_C^- ] \Big)
    \bigg] \ .
\end{align}

It is now simple to check that the symmetries of the Bosonic extension extend to symmetries of $S_F$. We see that it is invariant under the spatial transformations \eqref{eq: spatial symmetry} provided we take
\begin{subequations} \label{eq: fermion spatial}
\begin{align}
    \hat{\psi}_1^-(\hat{t}, \hat{z} , \hat{\bar{z}}) &= \brac{1 - \bar{\partial} \bar{f}} \psi_1^- (t,z,\bar{z}) \ , \\
    \hat{\psi}_1^+(\hat{t}, \hat{z} , \hat{\bar{z}}) &= \brac{\psi_1^+ - \inv{2} \bar{f}' \psi_1^-}(t,z,\bar{z}) \ , \\
    \hat{\psi}_A^-(\hat{t}, \hat{z} , \hat{\bar{z}}) &= \brac{\psi_A^- - \inv{2} f' \psi_A^+}(t,z,\bar{z}) \ , \\
    \hat{\psi}_A^+(\hat{t}, \hat{z} , \hat{\bar{z}}) &= \brac{1 - \partial f} \psi_A^+ (t,z,\bar{z}) \ ,
\end{align}
\end{subequations}
and the temporal transformations \eqref{eq: temporal symmetry} if
\begin{subequations}  \label{eq: fermion temporal}
\begin{align}
    \hat{\psi}_1^-(\hat{t} , \hat{z}, \hat{\bar{z}}) &= \psi_1^-(t,z,\bar{z}) \ , \\
    \hat{\psi}_1^+(\hat{t} , \hat{z}, \hat{\bar{z}}) &= \brac{1 - F'}\psi_1^+(t,z,\bar{z}) \ , \\
    \hat{\psi}_A^-(\hat{t} , \hat{z}, \hat{\bar{z}}) &= \brac{1 - F'} \psi_A^-(t,z,\bar{z}) \ , \\
    \hat{\psi}_A^+(\hat{t} , \hat{z}, \hat{\bar{z}}) &= \psi_A^+(t,z,\bar{z}) \ .
\end{align}
\end{subequations}
The transformations of the Fermions under the global symmetries are
\begin{equation} \label{eq: fermion SU(3)}
    \hat{\bpsi}^{A,\pm} = \tensor{\cal{R}}{^A_B} \bpsi^{B,\pm}
\end{equation}
for the $SU(3)$ R-symmetry,
\begin{subequations} \label{eq: fermion U(1)R}
\begin{align}
    \hat{\psi}_1^{\pm} &= e^{-\frac{i \alpha}{2}} \psi_1^{\pm} \ , \\
    \hat{\psi}_A^{\pm} &= e^{\frac{i \alpha}{2}} \psi_A^{\pm} \ ,
\end{align}
\end{subequations}
for the $U(1)_R$ R-symmetry, and
\begin{subequations} \label{eq: fermion U(1)b}
\begin{align}
    \hat{\psi}_1^{\pm} &= e^{i\beta} \psi_1^{\pm} \ , \\
    \hat{\psi}_A^{\pm} &= e^{i\beta} \psi_A^{\pm} \ ,
\end{align}
\end{subequations}
for the $U(1)_b$ baryon number symmetry.

\subsection{Fermionic Contributions to Conserved Currents} \label{sect: currents fermions}

In section \ref{sect: currents} we determined the contributions to the conserved currents in the theory solely coming from Bosonic fields- in this appendix we extend this to include the Fermionic terms.

The temporal transformations \eqref{eq: fermion temporal} give the current
\begin{subequations}
\begin{align}
    j^0_{(a)} &= \tr\brac{ 2i\brac{ \bpsi^{1,+} D \psi_1^- +  \bpsi^{1,-} \bar{D} \psi_1^+ +  \bpsi^{A,+} D\psi_A^- +  \bpsi^{A,-} \bar{D}\psi_A^+} + V_f } \ , \\
    j_{(a)} &= - 2i \tr\brac{ \bpsi^{1,+} D_0 \psi_1^- + \bpsi^{A,+} D_0 \psi_A^- } \ , \\
    \bar{j}_{(a)} &= - 2i \tr\brac{ \bpsi^{1,-} D_0 \psi_1^+ + \bpsi^{A,-} D_0 \psi_A^+ } \ ,
\end{align}
\end{subequations}
for $F=a$; it is then easy to see that other currents are just this multiplied by powers of $t$, {\it i.e.}
\begin{subequations}
\begin{align}
    j^{\mu}_{(b)} &= t j^{\mu}_{(a)} \ , \\
    j^{\mu}_{(c)} &= t^2 j^{\mu}_{(a)} \ .
\end{align}
\end{subequations}
The spatial transformations \eqref{eq: fermion spatial} give the holomorphic current
\begin{subequations}
\begin{align}
    T = 2i \tr \brac{D \bpsi^{A,-} \psi_A^+ - \bpsi^{1,-} D\psi_1^+} \ .
\end{align}
\end{subequations}
Finally, the R-symmetry transformations give the currents
\begin{subequations}
\begin{align}
    \tensor{\brac{J^0}}{^A_B} &= \tr \brac{\bpsi^{A,+} \psi_B^+} \ , \\
    \tensor{\brac{J}}{^A_B} &= \tr\brac{\bpsi^{A,+} \psi_B^-} \ , \\
    \tensor{\brac{\bar{J}}}{^A_B} &= \tr\brac{\bpsi^{A,-} \psi_B^+} \ , 
\end{align}
\end{subequations}
for the $SU(3)$ symmetry \eqref{eq: fermion SU(3)},
\begin{subequations}
\begin{align}
    j^0 &= \inv{2}\tr\brac{\bpsi^{1,-} \psi_1^- - \bpsi^{A,+} \psi_A^+} \ , \\
    j &= \tr\brac{\bpsi^{A,+} \psi_A^- - \bpsi^{1,+} \psi_1^-} \ , \\
    \bar{j} &= \tr\brac{\bpsi^{A,-} \psi_A^+ - \bpsi^{1,-} \psi_1^+} \ ,
\end{align}
\end{subequations}
for the $U(1)_R$ symmetry \eqref{eq: fermion U(1)R}, and the baryon number $U(1)_b$ symmetry \eqref{eq: fermion U(1)b} gives
\begin{subequations}
\begin{align}
    j_b^0 &= -\tr\brac{\bpsi^{1,-} \psi_1^- + \bpsi^{A,+} \psi_A^+} \ , \\
    j_b &= 2\tr\brac{\bpsi^{A,+} \psi_A^- + \bpsi^{1,+} \psi_1^-} \ , \\
    \bar{j_b} &= 2\tr\brac{\bpsi^{A,-} \psi_A^+ + \bpsi^{1,-} \psi_1^+} \ .
\end{align}
\end{subequations}

\section{Equations of Motion} \label{sect: eom}

The equations of motion that come from the actions \eqref{eq: bosonic action} and \eqref{eq: fermionic action} are
\begin{subequations} \label{eq: equations of motion}
\begin{align} \nonumber
    &- D_0^2 \calz^1 + \frac{2 \pi i}{k} \sbrac{\calz^1, D_0 \calz^A ; \bcalz_A} + \frac{2 \pi i}{k} \sbrac{\calz^A, \calz^1 ; D_0 \bcalz_A} + \frac{4\pi^2}{3k^2} \sbrac{ \calz^B, \sbrac{\calz^A, \calz^1 ; \bcalz_A } ; \bcalz_B} \\ \nonumber
    &- \frac{16\pi^2}{3k^2} \sbrac{\calz^B , \sbrac{\calz^A, \calz^1 ; \bcalz_B}; \bcalz_A } - \frac{8\pi^2}{3k^2} \sbrac{\calz^A, \calz^B ; \sbrac{ \bcalz_A, \bcalz_B ; \calz^1} } \\ \nonumber
    &- \frac{4\pi^2}{3k^2} \sbrac{\calz^A, \calz^1 ; \sbrac{\bcalz_B, \bcalz_A ; \calz^B }} - \frac{4\pi^2}{3k^2} \sbrac{\calz^1 , \sbrac{\calz^B, \calz^A ; \bcalz_B} ; \bcalz_A} - DH \\ \nonumber
    &+ \frac{4\pi}{k} \{\calz^1 , \psi_1^+ ; \bpsi^{1,+}\} + \frac{4\pi}{k} \{\calz^1 , \psi_A^- ; \bpsi^{A,-}\} + \frac{4\pi}{k} \{ \calz^A , \psi^+_A ; \bpsi^{1,+} \} \\
    &- \frac{4\pi}{k} \{\calz^A , \psi_A^- ; \bpsi^{1,-} \} - \frac{4\pi}{k} \epsilon^{ABC} \{\psi_A^+, \psi_C^- ; \bcalz_B \} = 0 \ ,
\end{align}
\begin{align} \nonumber
    &2( D \Bar{D} + \Bar{D} D) \calz^A - \frac{4\pi i}{k} \sbrac{\calz^1 , D_0 \calz^A ; \bcalz_1 } - \frac{2\pi i}{k} \sbrac{D_0 \calz^1, \calz^A; \bcalz_1} \\ \nonumber
    & - \frac{2\pi i}{k} \sbrac{\calz^1, \calz^A; D_0 \bcalz_1} + \frac{16 \pi^2}{3k^2} \sbrac{\calz^1 , \sbrac{\calz^A, \calz^B ; \bcalz_1}; \bcalz_B} \\ \nonumber
    & - \frac{4\pi^2}{3k^2} \sbrac{\calz^1, \calz^A ; \sbrac{\bcalz_B, \bcalz_1 ; \calz^B}} - \frac{4\pi^2}{3k^2} \sbrac{\calz^A , \sbrac{\calz^B, \calz^1 ; \bcalz_B}; \bcalz_1} \\ \nonumber
    &+ \frac{16\pi^2}{3k^2} \sbrac{\calz^B, \sbrac{\calz^A, \calz^1 ; \bcalz_B}; \bcalz_1} 
    + \frac{16\pi^2}{3k^2} \sbrac{\calz^1, \calz^B ; \sbrac{\bcalz_B, \bcalz_1 ; \calz^A}} \\ \nonumber
    &+ \frac{4\pi^2}{3k^2} \sbrac{\calz^B , \sbrac{\calz^1 , \calz^A; \bcalz_1} ; \bcalz_B} - \frac{4\pi^2}{3k^2} \sbrac{\calz^A, \sbrac{\calz^1, \calz^B ; \bcalz_1} ; \bcalz_B} \\ \nonumber
    &- \frac{4\pi^2}{3k^2} \sbrac{\calz^1, \sbrac{\calz^A, \calz^B ; \bcalz_B} ; \bcalz_1} + \frac{4\pi^2}{3k^2} \sbrac{\calz^A, \calz^B ; \sbrac{\bcalz_1, \bcalz_B ; \calz^1}} \\ \nonumber
    &+ \frac{2\pi}{k} \{ \calz^A , \psi_1^- ; \bpsi^{1,-} \} - \frac{2\pi}{k} \{ \calz^A , \psi_B^+ ; \bpsi^{B,+} \} + \frac{4\pi}{k} \{ \calz^1, \psi_1^+ ; \bpsi^{A,+} \} \\ 
    &- \frac{4\pi}{k} \{ \calz^1 , \psi_1^- ; \bpsi^{A,-} \} + \frac{4\pi}{k} \{ \calz^B , \psi_B^+ ; \bpsi^{A,+} \} + \frac{4\pi}{k} \epsilon^{ABC} \{ \psi_A^+ , \psi_C^- ; \bcalz_1 \}
    = 0 \ ,
\end{align}
\begin{equation}
    \Bar{D} \calz^1 = 0 \ ,
\end{equation}
\begin{align}
    F_{0z}^L - \frac{\pi}{k} \calz^1 \Bar{H} + \frac{2\pi}{k} \bigg( \calz^A D \bcalz_A - D\calz^A \bcalz_A \bigg) - \frac{2\pi i}{k} \bigg( \psi_1^+ \bpsi^{1,-} + \psi_A^+ \bpsi^{A,-} \bigg) &= 0 \ , \\
    F_{0z}^R - \frac{\pi}{k} \Bar{H} \calz^1 - \frac{2\pi}{k} \bigg( \bcalz_A D\calz^A - D \bcalz_A \calz^A \bigg) + \frac{2\pi i}{k} \bigg( \bpsi^{1,-} \psi_1^+ + \bpsi^{A,-} \psi_A^+ \bigg) &= 0 \ ,
\end{align}
\begin{align} \nonumber
    F_{z\Bar{z}}^L - \frac{\pi}{k} \bigg( \calz^1 D_0 \bcalz_1 - D_0 \calz^1 \bcalz_1\bigg) &- \frac{2\pi^2 i}{k^2} \bigg( \calz^A \sbrac{\bcalz_1, \bcalz_A; \calz^1} - \sbrac{\calz^1 , \calz^A ; \bcalz_1} \bcalz_A \bigg) \\
    &+ \frac{i \pi}{k} \bigg( \psi_1^- \bpsi^{1,-} + \psi_A^+ \bpsi^{A,+} \bigg) = 0 \ , \\ \nonumber
    F_{z\Bar{z}}^R + \frac{\pi}{k} \bigg(   \bcalz_1 D_0 \calz^1 - D_0 \bcalz_1 \calz^1 \bigg) &+ \frac{2\pi^2 i}{k^2} \bigg( \bcalz_A\sbrac{\calz^1 , \calz^A ; \bcalz_1} - \sbrac{\bcalz_1, \bcalz_A; \calz^1} \calz^A    \bigg) \\
    &- \frac{i \pi}{k} \bigg(\bpsi^{1,-} \psi_1^- + \bpsi^{A,+} \psi_A^+ \bigg) = 0  \ ,
\end{align}
\begin{align}
    i D_0 \psi_1^- - 2i \bar{D} \psi_1^+ + \frac{2\pi}{k} [\psi_1^-, \calz^A ; \bcalz_A] - \frac{4\pi}{k} [\psi_A^-, \calz^A ; \bcalz_1] &= 0 \ , \\
     D \psi_1^- + \frac{2\pi i}{k} [ \psi_1^+, \calz^1 ; \bcalz_1] + \frac{2 \pi i}{k} [ \psi^+_A , \calz^A ; \bcalz_1] &= 0 \ , \\ \nonumber
    i D_0 \psi_A^+ - 2i D \psi_A^- - \frac{2\pi}{k} [ \psi_A^+ , \calz^B ; \bcalz_B] + \frac{4\pi}{k} [\psi_1^+ , \calz^1 ; \bcalz_A]& \\
    + \frac{4\pi}{k} [\psi_B^+, \calz^B ; \bcalz_A] + \frac{4\pi}{k} \epsilon_{ABC} [\calz^B , \calz^1 ; \bpsi^{C,-}] &= 0 \ , \\
    \bar{D} \psi_A^+ + \frac{2\pi i}{k} [\psi_A^- , \calz^1 ; \bcalz_1] - \frac{2\pi i}{k} [\psi_1^- , \calz^1 ; \bcalz_A] - \frac{2\pi i}{k} \epsilon_{ABC} [\calz^B , \calz^1 ; \bpsi^{C,+}] &= 0 \ ,
\end{align}
\end{subequations}
where we've defined
\begin{equation}
    \{A,B ; \bar{C} \} = A \bar{C} B + B \bar{C} A \ .
\end{equation}
To find classical solutions we set all Fermions to zero. The obvious set of solutions to these are those with
\begin{subequations}
\begin{align}
    A_0^L = A_0^R &= 0 \ , \\
    A_z^L = A_z^R &\equiv A_z \ , \\
    \partial_0 A_z &= 0 \ , \\
    \partial_0 \calz^1 &= 0 \ , \\
    \calz^A &= w^A \bbm{1}_N \ , \\
    H &= 0 \ ,
\end{align}
\end{subequations}
with $v^A$ constant, so that the equations reduce to
\begin{subequations}
\begin{align}
    \bar{D} \calz^1 \equiv \bar{\partial} \calz^1 - i [A_{\bar{z}},\calz^1] &= 0 \ , \\
    F_{z\Bar{z}} + \frac{4\pi^2 i}{k^2} w^A \bar{w}_A [\calz^1 , \bcalz_1] &= 0 \ .
\end{align}
\end{subequations}
We recognise these as the Hitchin equations \cite{Hitchin:1986vp}. A more detailed analysis of the solutions, including time dependence was given in \cite{Kucharski:2017jwv} for the gauge group $SU(2)\times SU(2)$.

\section{Subleading Fields in the MNC Limit} \label{sect: 11d limit}

The MNC limit performed in \cite{Blair:2021waq} and \cite{Bergshoeff:2023igy} worked in the absence of subleading fields; in this appendix we show how these can be included, with the end result being that the dynamics of the theory are unaltered. We will first review the limit as it is presented in \cite{Bergshoeff:2023igy}. In terms of the hatted variables \eqref{eq: hatted variables} and 
\begin{equation}
    \hat{F}_4 = d\brac{C_3 + \inv{c^3} \Tilde{C}_3 + O\brac{\inv{c^6}}} \ ,
\end{equation}
the divergent piece of the action is
\begin{align}
    S_3 = - \inv{2 \cdot 4!} \int d^{11}x \, \hat{\Omega} \hat{F}_{\mu_1 \mu_2 \mu_3 \mu_4} \brac{ \hat{H}^{\mu_1 \nu_1} ... \hat{H}^{\mu_4 \nu_4} + \inv{4! 3! \hat{\Omega}} \hat{\varepsilon}^{\mu_1 ... \mu_4 \nu_1 ... \nu_7} \epsilon_{abc} \hat{\tau}_{\nu_5}^a \hat{\tau}_{\nu_6}^b \hat{\tau}_{\nu_7}^c } \hat{F}_{\nu_1 \nu_2 \nu_3 \nu_4} \ .
\end{align}
The bracketed terms can be rewritten as
\begin{equation}
    \inv{2} \brac{\hat{H}^{\mu_1 [\nu_1} ... \hat{H}^{|\mu_4| \nu_4]} + \inv{4! 3! \hat{\Omega}} \hat{\varepsilon}^{\mu_1 ... \mu_4 \nu_1 ... \nu_7} \epsilon_{abc} \hat{\tau}_{\nu_5}^a \hat{\tau}_{\nu_6}^b \hat{\tau}_{\nu_7}^c } = \hat{e}^{\mu_1}_{I_1} ... \hat{e}^{\mu_4}_{I_4} P_{I_1 ... I_4 J_1 ... J_4} \hat{e}^{\nu_1}_{J_1} ... \hat{e}^{\nu_4}_{J_4}\ ,
\end{equation}
in terms of the local $SO(8)$ tensor
\begin{equation}
    P_{I_1 ... I_4 J_1 ... J_4} = \inv{2} \brac{
    \delta^{[J_1}_{I_1} ... \delta^{J_4]}_{I_4} - \inv{4!} \epsilon_{I_1 ... I_4 J_1 ... J_4} }\ ,
\end{equation}
that projects local 4-forms onto their anti-self-dual part. Defining
\begin{equation}
    \hat{f}_{I_1 ... I_4} = P_{I_1 ... I_4 J_1 ... J_4} \hat{e}^{\mu_1}_{J_1} ... \hat{e}^{\mu_4}_{J_4} \hat{F}_{\mu_1 ... \mu_4} \ ,
\end{equation}
we see that the divergent part of the action is just
\begin{equation} \label{eq: hatted action}
    S_3 = - \inv{4!} \int d^{11}x \, \hat{\Omega} \, \hat{f}_{I_1 ... I_4} \hat{f}^{I_1 ... I_4} \ .
\end{equation}

The contributions of the subleading fields to the action can then be found by performing the $\inv{c^3}$ expansion of the hatted variables. This gives
\begin{align} \nonumber
    S_3 = - \inv{4!} \int d^{11}x \, \Omega \bigg[& f_{I_1 ... I_4} f^{I_1 ... I_4} + \frac{2}{c^3} \bigg( 4 \Phi^{\mu_1}_{I_1} e_{I_2}^{\mu_2} e_{I_3}^{\mu_3} e_{I_4}^{\mu_4} F_{\mu_1 \mu_2 \mu_3 \mu_4} + e^{\mu_1}_{I_1} ... e^{\mu_4}_{I_4} \Tilde{F}_{\mu_1 \mu_2 \mu_3 \mu_4} \\
    & + \inv{2} \brac{\frac{\partial \ln \Omega}{\partial \tau^a_{\mu}} m_{\mu}^a + \frac{\partial \ln \Omega}{\partial e^J_{\mu} } \pi^J_{\mu} } F_{I_1 ... I_4} 
    \bigg) f^{I_1 ... I_4} \bigg] + O\brac{\inv{c^6}} \ ,
\end{align}
where $f_{I_1 ... I_4}$ is the leading-order contribution to $\hat{f}_{I_1 ... I_4}$. As the divergent term is a squared quantity we can introduce a Hubbard-Stratonivch field $g_{I_1 ... I_4}$ to write it as
\begin{equation}
    -\frac{c^3}{4!} \int d^{11}x \, \Omega \, f_{I_1 ... I_4} f^{I_1 ... I_4} \to - \inv{4!} \int d^{11} x \, \Omega \brac{ 2 g_{I_1 ... I_4} f^{I_1 ... I_4} - \inv{c^3} g_{I_1 ... I_4} g^{I_1 ... I_4}} \ ,
\end{equation}
with the second form reducing to the first when $g_{I_1 ... I_4}$ satisfies its equation of motion
\begin{equation} \label{eq: g eom}
    g_{I_1 ... I_4} = c^3 f_{I_1 ... I_4} \ .
\end{equation}
The contribution from $S_3$ is then finite. Using $\cal{L}$ to denote the explicitly $O(1)$ part of the expanded Lagrangian and taking $c\to\infty$, we see that the action becomes
\begin{align} \nonumber
    S =  \int d^{11}x \, \Omega \bigg(& \cal{L} - \frac{2}{4!} \bigg[g_{I_1 I_2 I_3 I_4} +  4 \Phi^{\mu_1}_{I_1} e_{I_2}^{\mu_2} e_{I_3}^{\mu_3} e_{I_4}^{\mu_4} F_{\mu_1 \mu_2 \mu_3 \mu_4} +  \Tilde{F}_{I_1 I_2 I_3 I_4} \\ \nonumber
    & + \inv{2} \brac{\frac{\partial \ln \Omega}{\partial \tau^a_{\mu}} m_{\mu}^a + \frac{\partial \ln \Omega}{\partial e^J_{\mu} } \pi^J_{\mu} } F_{I_1 ... I_4} 
    \bigg] f^{I_1 ... I_4} \bigg) \\ \label{eq: regulated action}
    \equiv \int d^{11}x \, \Omega \bigg(& \cal{L} - \frac{2}{4!} G_{I_1 ... I_4} f^{I_1 ... I_4} \bigg)  \ .
\end{align}

It is straightforward to check that the equations of motion obtained from this action are equivalent to those obtained by treating $G_{I_1 ... I_4}$ as an independent field and substituting its explicit form into the equations. A quick way to see this must be the case is to note that we could have instead introduced a Hubbard-Stratonovich field for the unexpanded divergent term \eqref{eq: hatted action} to give a finite action; performing the expansion of the hatted variables then only gives $O\brac{\inv{c^3}}$ terms, and after taking $c\to\infty$ limit we find the second form of \eqref{eq: regulated action}, this time with $G_{I_1 ... I_4}$ as an independent field.

The only effect of the subleading fields is then to determine the on-shell value of the Lagrange multiplier $G_{I_1 ... I_4}$. In the $c\to \infty$ limit, the equation of motion \eqref{eq: g eom} along with the requirement that the field must be finite for any physical solution requires us to take
\begin{equation}
    f_{I_1 ... I_4} = O\brac{\inv{c^3}} \ .
\end{equation}
However, as we have already performed the $\inv{c^3}$ expansion for the fields $f_{I_1 ... I_4}$ can, by definition, only contain leading-order terms; this means we must have
\begin{equation}
    f_{I_1 ... I_4} = 0 \ ,
\end{equation}
which can also be seen by noting that in the limit $g_{I_1 ... I_4}$ becomes a Lagrange multiplier field. From the equation of motion we see that this implies
\begin{equation}
    g_{I_1 ... I_4} = 0\ ,
\end{equation}
also. The on-shell value of $G_{I_1 ... I_4}$ is then
\begin{equation} \label{eq: on-shell G}
    G_{I_1 ... I_4} = 4 \Phi^{\mu_1}_{I_1} e_{I_2}^{\mu_2} e_{I_3}^{\mu_3} e_{I_4}^{\mu_4} F_{\mu_1 \mu_2 \mu_3 \mu_4} +  \Tilde{F}_{I_1 I_2 I_3 I_4} + \inv{2} \brac{\frac{\partial \ln \Omega}{\partial \tau^a_{\mu}} m_{\mu}^a + \frac{\partial \ln \Omega}{\partial e^J_{\mu} } \pi^J_{\mu} } F_{I_1 ... I_4} \ ,
\end{equation}
which is entirely determined by the subleading fields.

\section{Non-Relativistic Orbifold Geometry} \label{sect: orbifold}

Recall that the dual of ABJM is the near-horizon limit of M2-branes on the background with transverse space $\bb{C}^4/\bb{Z}_k$. To begin we consider $\bb{C}^4$ with coordinates $z^M$:
\begin{align}
	 ds^2_{\bb{C}^4}  = dz^1d\bar z_1+dz^2d\bar z_2+dz^3d\bar z_3+dz^4d\bar z_4\ .
\end{align}
 In order to implement the orbifold we note that $S^{2n+1}$ can be realised as a circle fibration over $\bb{CP}^{2n}$. If we introduce the parameterisation
\begin{equation}
    z^M = r_M e^{i \theta_M} \ ,
\end{equation}
then the metric takes the form
\begin{equation}
    ds^2_{\bb{C}^4} = d\tilde r^2 + \tilde r^2 \brac{ds^2_{\bb{CP}^3} + \brac{d\tilde \phi + \tilde {\omega}}^2 } \ ,
\end{equation}
where $\tilde r = \sqrt{r_1^2+...+r_4^2}$, $\tilde \phi=\theta_1+...+\theta_4$  and $d {\tilde \omega}$ is the K\"ahler form on $\bb{CP}^3$. The orbifold then takes
\begin{equation}
    \tilde \phi \to \frac{\tilde \phi}{k} \ ,
\end{equation}
where we have retained the periodicity of $\tilde \phi\sim \tilde \phi+2\pi$. 

In order to take the non-relativistic limit we must make a distinction between $z^1$ and the other coordinates. It is therefore convenient to write the metric on  $\bb{C}^4$ as
\begin{equation}
    ds_{\bb{C}^4}^2 = dr_1^2 + r_1^2 d\theta^2_1 + dr^2 + r^2 \brac{ds_{\bb{CP}^2}^2 + \brac{d\phi + \omega}^2 } \ ,
\end{equation}
where $r = \sqrt{r_2^2 + r_3^2 + r_4^2}$, $\phi = \theta_2+\theta_3+\theta_4$,$ $  and $d\omega$ is the K\"ahler form on $\bb{CP}^2$. We note that this means that
\begin{equation}
    \tilde \phi = \phi + \theta_1 \ .
\end{equation}
We can then implement the non-relativistic limit of the metric, obtaining the (partial) Newton-Cartan structures
\begin{subequations}
\begin{align}
    \hat{P}_{\mn} dx^{\mu}\otimes  dx^{\nu} =& \,dr_1\otimes d  r_1+ r_1^2d\theta_1\otimes d\theta_1\ , \\ \nonumber
    \hat{Q}^{\mn} \frac{\partial}{\partial x^{\mu}} \otimes \frac{\partial}{\partial x^{\nu}} =& \, \frac{\partial}{\partial r} \otimes \frac{\partial}{\partial r} + \inv{r^2} \Bigg( g^{mn}_{(\bb{CP}^2)} \frac{\partial}{\partial z^m} \otimes \frac{\partial}{\partial  z^n} - \omega^i \brac{ \frac{\partial}{\partial z^m} \otimes \frac{\partial}{\partial \phi} + \frac{\partial}{\partial \phi} \otimes \frac{\partial}{\partial z^m} } \\
    &+ \brac{1 + \omega^i \omega_i} \frac{\partial}{\partial \phi} \otimes \frac{\partial}{\partial \phi} \Bigg) \ ,
\end{align}
\end{subequations}
where $\{z^m\}$ are some local coordinates on $\bb{CP}^2$ and we've defined $\omega^m = g^{mn}_{(\bb{CP}^2)} \omega_{n}$. We see that if we make the angular coordinate transformation $(\theta_1,\phi)\to(\theta_1,\tilde \phi)$ the form of both structures is unchanged, {\it i.e.} we just replace $\phi$ with $\tilde\phi$ in the expression for $\hat{Q}$. We can then implement the orbifold, leaving us with
\begin{subequations}
\begin{align}
    P_{\mn} dx^{\mu}\otimes  dx^{\nu} =& \,dr_1\otimes d  r_1+ r_1^2d\theta_1\otimes d\theta_1 \ , \\ \nonumber
    Q^{\mn} \frac{\partial}{\partial x^{\mu}} \otimes \frac{\partial}{\partial x^{\nu}} =& \, \frac{\partial}{\partial r} \otimes \frac{\partial}{\partial r} + \inv{r^2} \Bigg( g^{mn}_{(\bb{CP}^2)} \frac{\partial}{\partial z^m} \otimes \frac{\partial}{\partial z^n} - k \omega^m \brac{ \frac{\partial}{\partial z^m} \otimes \frac{\partial}{\partial \tilde \phi} + \frac{\partial}{\partial \tilde \phi} \otimes \frac{\partial}{\partial z^m} } \\
    &+ k^2\brac{1 + \omega^m \omega_m} \frac{\partial}{\partial \tilde\phi} \otimes \frac{\partial}{\partial \tilde \phi} \Bigg) \ .
\end{align}
\end{subequations}
This defines a non-relativistic $\bb{C} \times \bb{C}^3/\bb{Z}_k$   geometry, with the orbifold only acting on the co-metric.

\printbibliography
 
\end{document}